\documentclass{aa}  

\usepackage{graphicx}
\usepackage{txfonts}
\usepackage{cellspace} 
\usepackage{xcolor}
\usepackage{placeins}
\usepackage{natbib,twoopt}
\usepackage[breaklinks=true]{hyperref} 
\bibpunct{(}{)}{;}{a}{}{,}             
\makeatletter
  \newcommandtwoopt{\citeads}[3][][]{\href{http://adsabs.harvard.edu/abs/#3}%
    {\def\hyper@linkstart##1##2{}%
     \let\hyper@linkend\@empty\citealp[#1][#2]{#3}}}
  \newcommandtwoopt{\citepads}[3][][]{\href{http://adsabs.harvard.edu/abs/#3}%
    {\def\hyper@linkstart##1##2{}%
     \let\hyper@linkend\@empty\citep[#1][#2]{#3}}}
  \newcommandtwoopt{\citetads}[3][][]{\href{http://adsabs.harvard.edu/abs/#3}%
    {\def\hyper@linkstart##1##2{}%
     \let\hyper@linkend\@empty\citet[#1][#2]{#3}}}
  \newcommandtwoopt{\citeyearads}[3][][]%
    {\href{http://adsabs.harvard.edu/abs/#3}
    {\def\hyper@linkstart##1##2{}%
     \let\hyper@linkend\@empty\citeyear[#1][#2]{#3}}}
\makeatother

\newcommand{\kms}{km\,s\ensuremath{^{-1}}}

\newcommand{\teff}{\ensuremath{T_\text{eff}}}
\newcommand{\logg}{\ensuremath{\log g}}

\newcommand{\msun}{M\ensuremath{_\sun}}
\newcommand{\rsun}{R\ensuremath{_\sun}}

\newcommand{\betaur}{\ensuremath{\beta}~Aur}
\newcommand{\ellc}{\texttt{ellc}}
\newcommand{\orbitize}{\texttt{orbitize!}}
\newcommand{\oimodeler}{\texttt{oimodeler}}

\begin{document} 

   \title{Interferometric Survey of Stellar Parameters:\\ Mass of the metallic A-type binary $\beta$ Aur }

   \author{J.~Jonák \inst{1}
            \and D.~Mourard \inst{1}
            \and J.~D.~Monnier \inst {2}
            \and S.~Hoin \inst{2}
            \and M.~Brož \inst{3}
            \and A.~Oplištilová \inst{4}
            \and K.~A.~Kubiak \inst{5}
            \and N.~Ebrahimkutty \inst{1} 
            \and R.~V.~Ibañez-Bustos \inst{1} 
            \and H.~Nowacki \inst{1} 
            \and M.~Vrard  \inst{1}
            \and M.~Bailleul \inst{1}
            \and P.~Bério    \inst{1}
            \and J.~Dejonghe \inst{1}
            \and R.~Ligi \inst{1}
            \and F.~Morand \inst{1}
            \and N.~Nardetto \inst{1}
            \and D.~Salabert \inst{1}
            \and S.~Deheuvels    \inst{6}
            \and A.~Domiciano~de~Souza       \inst{1}
            \and A.~Meilland       \inst{1}
            \and K.~Perraut       \inst{7}
            \and M.~Wittkowski    \inst{8}
            \and S.~Kraus \inst{9}
            \and N.~Anugu \inst{5}
            \and M.~Gutierrez \inst{2}
        }

   \institute{Université Côte d’Azur, Observatoire de la Côte d’Azur, CNRS, Laboratoire Lagrange, Bd de l’Observatoire, 06304 Nice, France \\
                \email{juraj.jonak@oca.eu}
             \and
             Astronomy Department, University of Michigan, Ann Arbor, MI 48109, USA
             \and
             Charles University, Faculty of Mathematics and Physics, Astronomical Institute, V Holešovičkách 2, 180\,00 Praha 8, Czech Republic
             \and 
             GAPHE, STAR, Université de Liège, B5c, Allée du 6 Août 19c, B-4000 Sart Tilman, Liège, Belgium
             \and 
             The CHARA Array of Georgia State University, Mount Wilson Observatory, Mount           Wilson, CA 91203, USA
             \and 
             IRAP, Université de Toulouse, CNRS, CNES, UPS, 14 Avenue Edouard Belin, 31400 Toulouse, France
             \and
             Univ. Grenoble Alpes, CNRS, IPAG, 38000 Grenoble, France
             \and 
             European Southern Observatory, Karl-Schwarzschild-Str. 2, 85748 Garching bei München, Germany
             \and 
             Astrophysics Group, Department of Physics \& Astronomy, University of Exeter, Stocker Road, Exeter, EX4 4QL, UK
        }
   \date{Received May 22, 2026; accepted June 22, 2026}
 
  \abstract 
   {Long-baseline optical interferometry provides spatially resolved observations of close binaries, complementing spectroscopic and photometric constraints on stellar parameters.}
   {With the capabilities of the new visible CHARA/SPICA instrument and the multiple spectral band operation of CHARA, our goal is to resolve orbits of short-period binaries and develop a robust framework for combining interferometric, spectroscopic, and photometric observations into a single consistent model. }
   {For our target sample, we selected suitable binaries based on brightness, angular separation, and orbital properties based on the expected performance of the CHARA/SPICA instrument. As a case study, we analysed the bright eclipsing binary $\beta$ Aurigae, composed of two slightly evolved A1 stars. We combined new interferometric observations of \betaur~obtained with CHARA/SPICA, MIRC--X, and MYSTIC with archival MIRC data, radial velocities, and light curves. We first derived astrometric positions from interferometric observables and computed an orbital solution. Afterwards, we implemented a unified model, capable of tying interferometric modelling with the \texttt{ellc} code to simultaneously fit all observables using MCMC sampling. We performed a detailed analysis of the noise statistics of each data set and in the end we adopted a profile likelihood approach to account for underestimated noise and systematics.}
   {We derived a consistent orbital and physical solution for \betaur~through joint modelling. The inclusion of interferometric data tightly constrains the angular semi-major axis and inclination. Using profile likelihood to account for the different intrinsic levels of uncertainty of the fundamentally different observables, we derived the masses of the two stars, $M_1 = 2.359 \pm 0.005$\,\msun and $M_2 = 2.293 \pm 0.004$\,\msun, their radii $R_1 = 2.752 \pm 0.002$\,\rsun~and $R_2 = 2.622 \pm 0.002$\,\rsun, and the distance to the binary, $d = 24.30\pm0.05$\,pc.} 
   {}


   \maketitle
   \nolinenumbers 

\section{Introduction}

The Interferometric Survey of Stellar Parameters (ISSP) \citep{SPICA} aims to address key questions about the relation between planets and stars and to offer to the broader community a unique and primary source of direct information all over the Hertzsprung–-Russell (HR) diagram, and is discussed in \cite{spica2026} (hereafter Paper 1). A specific part is dedicated to determining dynamic masses of spectroscopic binaries and the results can be utilised by the other sub-programmes.

The mass of a star is a fundamental parameter driving the stellar structure and evolution as well as its terminal stage. Many different methods of determining it exist with varying accuracy, precision and model-dependence. \citet{Serenelli2021} present a comprehensive mass ladder, where, at the very base, one finds spectroscopic eclipsing or visual binaries. These cover the entire range of stellar masses with an accuracy well below 1\% and provide the only model-independent information on the stellar mass of the two components from the mutual interaction via radial velocity (RV) of orbital motion. 

The spectroscopic elements were derived by measuring the radial velocities (RVs) of the two components through the Doppler shift of spectral lines that originate from both stars, which follow a curve in time of
\begin{equation}
    RV_n(t) = \gamma - (-1)^n K_n \left[ \cos(\omega + \varv(t)) + e \cos\omega\right].
\end{equation}
Here $n \in \{1,2\}$ denotes the component, $\gamma$ the systemic velocity, $K_n$ velocity semi-amplitudes, $e$ the eccentricity, $\omega$ the argument of periastron, and $\varv(t)$ the true anomaly, computed from the orbital period, $P$, and the periastron passage, $T_\textrm{peri}$. These elements are then directly related, by second Kepler's law, to the projected semi-major axis
\begin{equation}
    a \sin i = \frac{P}{2\pi}(K_1+K_2)\sqrt{1-e^2},
\end{equation}
and the projected masses
\begin{equation}
    M_n \sin^3 i = \frac{P}{2 \pi G}\frac{K_1 K_2}{K_n} (K_1+K_2)^2 (1-e^2)^\frac{3}{2}.
\end{equation}
The factor $\sin^3 i$ comes from a geometrical inclination angle, $i$, of the orbit to the line of sight. 
Although being an important dependence for mass determination, it cannot be solved through spectroscopy alone and as such, complementary observations are needed.

The situation is much improved if the object is an eclipsing binary, where the orbital inclination is restrained close to 90\,\degr. These provide the full information of the system through simultaneous modelling of the RV curve and the light curve. From the shape of eclipses with respect to time, one can determine the radii of the components, while their depth decouples the brightness of the individual stars. Thus it is possible to obtain precise values of temperature, luminosity, and distance. 
However, these are much rarer, as they require a small window of inclination angles to be viewed from. 

A different approach is utilising interferometric measurements. Although these do not require one to view the system edge-on, in order to satisfyingly resolve the two stars, they pose different constraints, such as similar brightness of the components and suitable separation. 
Furthermore, the combination of angular (from interferometry) and absolute separation (from spectroscopy) of the components directly provides the distance of the system. Additionally, using models combining all the different types of observations together can significantly improve our understanding of the system.  

Indeed, numerous works have been able to combine interferometric measurements with RVs, and reaching sub-percent precision on mass and distance determination. Some authors prefer to first model the relative positions and, as a next step, obtain the masses from combined RV + astrometry models; for example, \citet{Gallenne2016} (0.05\% mass precision), \citet{Gallenne2019} (0.04\%), and \citet{Lester2019A,Lester2019B} (both 0.3\%). In recent years, precise models using directly interferometric observables have emerged -- for example, those of \citet{Morales2022} (1.5\%) and \citet{Danner2025} (0.18\%) -- and this method has been used for more complicated systems, such as \citet{PYTERPOL}. 

Furthermore, the upcoming release of Gaia DR4 is expected to contain epoch astrometry of binaries via the motion of their photocentres. A single piece of interferometric data will permit us to alleviate the remaining flux-separation degeneracy in Gaia  binaries \citep{Kraus_2022}.

In this article, we analyse the very bright ($m_V = 1.9$) eclipsing binary \betaur, consisting of two slightly evolved A1\,IV stars that have very similar properties and orbit each other with a period of about 3.96\,d. The system has been studied since the end of the 19th century, with spectroscopic observations spanning more than a hundred years. Our objective is to compile and combine the available historic spectroscopic and photometric data, present our interferometric observations, and derive the parameters of this binary using a self-consistent model that can be applied to further targets of the programme to homogeneously and precisely derive the respective stellar parameters. 

Throughout the paper we use the following dates related to the Julian date ($\textrm{JD}$) -- a modified Julian date,
\begin{equation}
    \textrm{MJD} = \textrm{JD} - 2\,400\,000.5,
\end{equation}
and a heliocentric Julian date,
\begin{equation}
    \textrm{HJD} = \textrm{JD} + \textrm{HC}(\textrm{JD}, \textrm{RA},\textrm{DEC}) - 2\,400\,000,
\end{equation}
where $\textrm{HC}$ is the heliocentric correction based on the position of the star and the Earth at the time. Interferometric observations are reported in MJD, and all dates are converted to HJD in the combined models.

The structure of the article is as follows. In Section \ref{sec:targets}, the target selection for the programme is discussed. Section \ref{sec:betaur} introduces the binary \betaur~along with the adopted and performed observations. Section \ref{sec:interferometry} describes the interferometric modelling and the orbit retrieved. Section \ref{sec:robust} introduces a robust model combining interferometry, RVs, and light curves into a single model.

\section{Target selection}
\label{sec:targets}
Based on the expected capabilities of the CHARA/SPICA instrument to successfully resolve the individual orbits, we chose our targets based on the following criteria:
\begin{itemize}
    \item an observability of $\text{DEC} > -30 \degr$,
    \item a sufficient brightness of $m_V < 8$, 
    \item a relative flux ratio of $f = \frac{L_2}{L_1} > 0.06$,
    \item an angular semi-major axis of $\alpha \in [0.15, 10]$\,mas,
    \item an orbital period of $P < 1$\,yr,
    \item a system sufficiently detached, based on the estimation of the ratio of the angular semi-major axis, $\alpha$, and the angular diameter of the primary, $\theta_1$, $\alpha / \theta_1 > 2.5$.
\end{itemize}

\subsection{Eclipsing binaries}
We searched through four existing catalogues of eclipsing binaries, \citet{Torres10}, \citet{Eker14}, \citet{Southworth15}, and \citet{Graczyk19}, and, when necessary, estimated the relevant missing parameters; for example, computing the flux ratio from listed temperatures or estimating the angular diameter from the listed radius and distance or parallax. Taking into account the redundancy of objects within these catalogues, we found 36 objects suitable for our SPICA observations.

\subsection{Catalogue of spectroscopic binaries SB9}
\citet{Pourbaix04} published the ninth iteration of the catalogues of spectroscopic binary orbits. As of recently, the catalogue contains about 4000 orbits, their period, and velocity semi-amplitudes. 

As the catalogue contains different types of objects, such as Be stars or ellipsoidal variables, we automatically ran the objects through a simbad query to remove any targets not primarily classified as either a spectroscopic or an eclipsing binary. As a next step, we applied the conditions on visual magnitude, period, and declination. Furthermore, we limited the selection to only objects for which the secondary had a defined RV curve. By cross-checking with the previous catalogues and eliminating the candidates already checked, about 200 targets remained.

The absence of eclipses in these systems directly results in much less information being available. We then manually checked to find published articles on all of these systems so as to adopt the parameters. Furthermore, for binaries with components classified to be on the main sequence, we calculated a crude estimate of the primary star's properties using the equations by \citet{Harmanec1988}, which relate the effective temperature (roughly estimated from spectral type) to stellar mass and radius based on observed eclipsing binaries.

\subsection{The target sample}

\begin{figure}
    \centering
    \includegraphics[width=.85\linewidth]{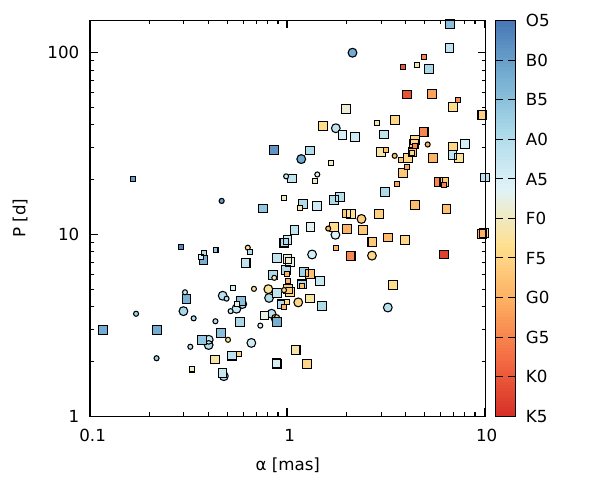}
    \includegraphics[width=.85\linewidth]{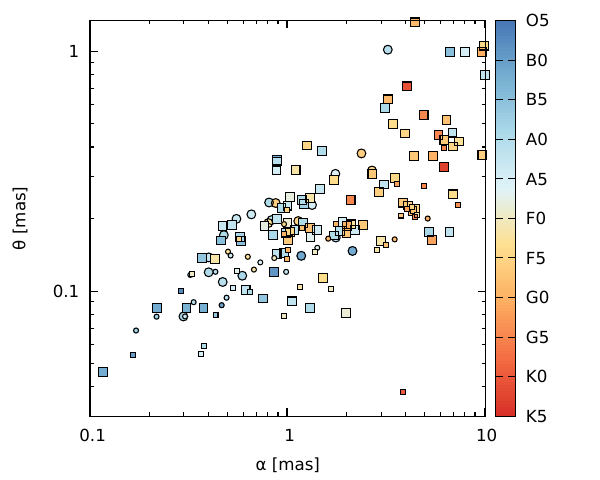}
    \caption{Distribution of (top) period, $P$, and angular semi-major axes, $\alpha$, and (bottom) estimated angular diameters, $\theta$ and $\alpha$, of our targets. The~objects are colour-coded by the respective spectral type of the primary, with circles representing eclipsing binaries and squares spectroscopic binaries. Furthermore, brighter stars $m_V < 7$ are plotted with larger symbols.}
    \label{fig:binaries_selection}
\end{figure}

The final sample contains, in total, 161 spectroscopic and eclipsing binaries. The distribution of periods, angular semi-major axes, and angular diameters can be seen in Fig. \ref{fig:binaries_selection}.

In this paper, we present the detailed methods applied on one specific target, \betaur, and the corresponding results. The results on the whole sample will be published in a further work. The entire sample of targets is available as online material.

We consider \betaur~to be a strong benchmark for validating our approach of combined modelling of photometry, spectroscopy, and interferometry. Works detailing RV observations span over a century, along with light curves of extremely high precision. This allows for both a test of the methodology presented further as well as a direct comparison with previous works before applying it to a broader sample of interferometric binaries.

\subsection{Complementary spectroscopy and photometry}
Due to the large number of potential targets, we assigned them priority based on already available spectroscopic data, which were necessary to complement our interferometric observations. To retrieve these, we created a code connecting our target database with a publicly available databases of observatories, such as Observatory Haute-Provence (OHP), Canada-France-Hawaii Telescope (CFHT) through Canadian Astronomy Data Centre (CADC), and the archive of the European Southern Observatory (ESO).

\section{Observations of \betaur}
\label{sec:betaur}
\subsection{Previous work}
Thanks to the brightness of \betaur~(HD\,40183), the history of its observations date to the first spectrometers \citep{Pickering1891,Maury1898,Vogel1904}, it being one of the first spectroscopic binaries discovered. The early observations were thoroughly examined by \citet{Baker1910}. Shortly after, \citet{Stebins1911} discovered eclipses of the system, making it one of the first eclipsing binaries as well. \citet{Smith1948} published a set of 21 RV measurements of both the primary and secondary components, observed with the 208cm McDonald telescope in 1943, with an expected accuracy of 0.7 \kms.

Its brightness actually proved to be a disadvantage, it being too bright for many telescopes, and as such only a few observations were made. \citet{Johansen1971} was able to obtain a light curve in eight passbands. These results were later used by \citet{Nordstrom1994} to derive the physical size of the components. In addition, \citet{Hummel1995} resolved the orbit using the Mark~III interferometer, noting the challenge of interferometrically observing the star due to its small angular separation (about 3\,mas) and short period. Nonetheless, their results were consistent with previous results and models.

\citet{Southworth2007} analysed an extremely rich light curve obtained by the WIRE satellite to calculate a robust solution, comparing the effects of different limb-darkening laws used in the model. Based on these high-precision data, they discovered a small but significant eccentricity that had previously gone unreported.  \citet{Behr2011} published 20 measurements of both primary and secondary RVs, by using a dispersed Fourier Transform Spectrograph. \citet{Strassmeier2020} revised the orbital solution by combining a continuous photometry with two BRITE satellites and high-resolution optical spectroscopy with STELLA.

\subsection{Adopting data}
We compiled multiple sources of previous spectroscopic observations, namely those published by \citet{Baker1910}, \citet{Smith1948}, \citet{Behr2011}, and \citet{Strassmeier2020}, as well as the photometry by \citet{Southworth2007} and \citet{Strassmeier2020}. The journal of adopted photometric and spectroscopic data is listed in Table \ref{tab:compdata}. 

Furthermore, within our search for complementary data, we found 18 publicly available spectra on the PolarBase database \citep{Polarbase1,PolarBase2}. These cover a large spectral range, from 360 to 880\,nm, with a resolution of 65\,000. All spectra were normalised with respect to the stellar continuum utilising the program \texttt{reSPEFO} \citep{reSPEFO1}. 
We extracted the RVs using the TODCOR algorithm \citep{TODCOR1,TODCOR2}, whereby observed spectra are compared to synthetic ones using a correlation function. As a base for creating the synthetic spectra, we used the ATLAS12 models from the POLLUX synthetic spectra database \citep{POLLUX}. Within this grid, we interpolated the spectra to the expected values of \teff\ and \logg\ of the two components, using the code \texttt{PYTERPOL} \citep{PYTERPOL}. The measured RVs are listed in Table~\ref{tab:narval}.

\begin{table}[]
    \centering\caption{Adopted photometric and spectroscopic observations.}
    \begin{tabular}{lcccc}\hline\hline
    \multicolumn{5}{c}{Photometry} \\
    Ref.    & Filter & \multicolumn{2}{c}{HJD range} &  \#  \\\hline
    1  & WIRE    & 53825      & 53846       & 4804   \\\hline
    2  & BRITE red  & 57655 & 57819    & 708   \\
       & BRITE blue & 57647 & 57822 & 1824  \\\hline\hline
    \multicolumn{5}{c}{Spectroscopy} \\
    Ref.    & $R$ & \multicolumn{2}{c}{HJD range} &  \#  \\\hline
    3  & 5\,000  &   18180 & 18344 &  46   \\
    4  & 40\,000 &   31047 & 31075 &  21 \\
    5  & 50\,000 &   54400 & 54491 &  20   \\
    2  & 55\,000 &   57487 & 57845 & 123 \\ \hline 
    \end{tabular}
    \tablefoot{First column corresponds to the source of the data. For photometry, we list the passband filter used, whilst for spectroscopy the resolving power $R$.}
    \tablebib{(1)~\citet{Southworth2007}; (2)~\citet{Strassmeier2020}; (3)~\citet{Baker1910}; (4)~\citet{Smith1948}; (5)~\citet{Behr2011}.}
    \label{tab:compdata}
\end{table}

\subsection{CHARA/SPICA, MIRC--X and MYSTIC observations}
We observed \betaur~ten times over eight nights between October 2023 and 2025. Most observations were performed simultaneously in a six-telescope configuration of CHARA in the $R$, $H$, and $K$ bands using Stellar Parameters and Images with a Cophased Array (SPICA; Paper 1), the Michigan InfraRed Combiner-eXeter (MIRC--X;  \citet{MIRCX}), and the Michigan Young STar Imager (MYSTIC; \citet{MYSTIC}).

The observations with SPICA were conducted in survey mode, focusing on a narrow strip in declination to optimise the movement of the telescopes as well as the possibility of sharing a calibrator star with the science targets of the different ISSP programmes. Our Night Scheduling Software selects stars within this strip as well as the most relevant calibrator stars. 

In some cases, due to uncooperative weather or technical issues, not all three instruments were used, or some observations were made with only five telescopes. The detailed log, along with the calibrators used for each night, can be found in Table \ref{tab:obslog}. The calibrators are stars with a small but well-known angular diameter (precision typically $<2\%$), and by comparing the measured (raw) visibilities to these expected values one can calculate the transfer function, which has information on the actual instrument and the atmospheric effect on the measured quantities. 

For the three instruments, the calibration step was done with the SPICA post-processing tools. A detailed description of the reduction and calibration procedures can be found in Paper~1, section 5. In general, the MIRC--X instrument was also used for tracking of the fringes. 

The reduction of the MIRC--X and MYSTIC data showed unexpectedly high noise in the result, especially for the calibrator stars. In order to fix this issue, special care had to be taken by increasing the standard thresholds of the flux and the signal-to-noise ratio. Furthermore, the difference in brightness between the  binary and the calibrators possibly deteriorated the stability of the transfer function.

Despite our best efforts performing the data reduction, the closure phases coming from SPICA were of very poor quality, unreliable for modelling, and we rejected them from further modelling.  Additionally, we put a strict minimum limit on the reported uncertainty of each observable, ensuring that $\sigma_{V^2} \ge 0.05 V^2$ and $\sigma_{\phi} \ge 1\degr$ for the respective data. 

\subsection{MIRC archival data}
To further improve the orbital coverage, we adopted ten additional observations that we retrieved from the CHARA~Database\footnote{\url{https://www.chara.gsu.edu/observers/database}}. These were taken using the CHARA/MIRC interferometer \citep{MIRC} between 2007 and 2016. We reduced these data following the manual\footnote{\url{https://chara.gsu.edu/tutorials/mirc-data-reduction}}.

\subsection{Wavelength recalibration}
\label{subsec:mm_miscal}

In addition, the current MIRC--X/MYSTIC pipeline does not account for a miscalibration of wavelength affecting the instrument. As noted by the MIRC-X/MYSTIC Pipeline Manual\footnote{v0.9.6, May 24 2024}, based on a comparison with results obtained with the VLTI/GRAVITY instrument \citep{GRAVITY},  we corrected this effect by dividing the measured wavelength by an instrument-dependent constant factor (i.e. $\lambda_\mathrm{true} = \lambda_\mathrm{measured} / \kappa$):
\begin{equation}
\begin{split}
    \kappa_\textrm{MIRC--X} &= {1.0054 \pm 0.0006}, \\
    \kappa_\textrm{MIRC--X} &= {0.999 \pm 0.001} \textrm{ (year 2025)}, \\
    \kappa_\textrm{MYSTIC} &= {1.0067 \pm 0.0007},\\
    \kappa_\textrm{MIRC} &= {1.0014 \pm 0.0006}.
\end{split}
\label{eq:mm_miscal}
\end{equation}

\section{Interferometric orbit}
\label{sec:interferometry}
As a first step in our modelling, we treated all the interferometric observations separately, with the goal of obtaining the relative position of the secondary for each epoch ($x(t), y(t)$). From these values we found a best-description orbit, as an intermediate step, to verify the internal consistency of these measurements as well as to derive the interferometry-specific parameters.

\subsection{From visibilities to positions}
An optical interferometer provides several observable quantities, all tied to the complex visibility, $\mu$, at spatial frequencies ($u,v$), which is, in turn, related to the Fourier transform of spatial distribution of the incoming intensity \citep{vanCittert1934,Zernike1938}. The two most commonly used quantities are the squared visibility,
\begin{equation}
    V^2(u,v) = |\mu(u,v)|^2,
\end{equation}
and closure phase (argument of the triple product on a closed triplet of baselines),
\begin{equation}
    \phi = \arg[\mu(u_1,v_1) \mu(u_2,v_2) \mu(-u_1-u_2,-v_1-v_2)].
\end{equation}
Assuming that the two stars of the system can be approximated as disks with quadratic limb-darkening (as introduced by \citet{Kopal1950}), with two coefficients, $A$ and $B$,
\begin{equation}
    \frac{I(\cos\vartheta)}{I(1)} = 1 - A(1-\cos\vartheta) - B(1-\cos\vartheta)^2,
\end{equation}
based on the angle of incidence, $\vartheta$, each component contributes to the complex visibility by
\begin{equation}
    \mu_n(u,v) = c\left[(1-A-B)\frac{J_1(t)}{t} + 
    \sqrt{\frac{\pi}{2}}(A+2B)\frac{J_{3/2}(t)}{t^{3/2}} \\
     - 2B\frac{J_2(t)}{t^2}\right],
\end{equation}
where the argument is related to the angular diameter, $\theta$, and wavelength, $\lambda$,
\begin{equation}
\label{eq:arg_ud}
t = \frac{\pi \sqrt{u^2+v^2}}{\lambda} \theta,
\end{equation}
and the normalisation constant, $c = \frac{12}{6-2A-B}$.

We adopted the values of these coefficients in the $R$, $H$, and $K$ bands from the table published by \citet{Claret2011}, taking the closest value in \teff~and \logg~to the expected value of our objects. As the properties of the two components are very similar, the same values of coefficients are used to describe the respective components. From the work of \citet{Southworth2007}, $\teff =$ 9350\,K and 9200\,K, and $\logg = $ 3.932 and 3.979, for the primary and secondary component, respectively, the closest values within the table of coefficients correspond to $\teff = 9250$\,K and $\logg = 4$.  The values of these coefficients are listed in Table~\ref{tab:LDcoeffs}.

Due to the linearity of the Fourier transform, each of the components gains a phase shift depending on its position $(x_n,y_n)$. The total complex visibility is 
\begin{equation}
\label{eq:complex_vis_binary}
    \mu(u,v) = \sum_{n=1,2} w_n \mu_n \exp\left[\frac{-2 \pi i}{\lambda} (ux_n+vy_n)\right],
\end{equation}
where $w_n$ is the normalised flux weight of each component,
\begin{equation}
   w_n = \frac{L_n}{L_1 + L_2},
   \label{eq:flux_weight}
\end{equation}
with $w_1 + w_2 = 1$. Alternatively, expressed from the flux ratio, $w_1 = 1/(1+f)$ and $w_2 = f/(1+f)$. However, the flux ratio of these two stars is not the same across the three spectral bands ($R$, $H$, and $K$ for SPICA, MIRC--X, and MYSTIC, respectively). We therefore introduced flux weights $w_R$, $w_H$, and $w_K$ in each band (as in Eq. \ref{eq:flux_weight}), and assumed that it is constant within the passband. 

In order to make an initial guess of the relative position of each observation, we used LITPro \citep{LitPRO}, whereby we ran a simple monochromatic model, using only the MIRC--X data. Afterwards, we calculated the posterior of the optimised parameters using the program \oimodeler~\citep{OIMODELER} through MCMC (implemented via the \texttt{emcee} package \citet{emcee}). For each observation we used 20 walkers for 10\,000 steps, and we varied up to seven free parameters: angular diameters, $\theta_n$, the position of the secondary $(x,y)$, and flux weights, $w_R$, $w_H$, and $w_K$ (if all instruments were used). For the positions, we set a uniform prior within 2\,mas of the solution from LITPro.

However, we found that the data from SPICA were not constraining the flux weight in the R band, giving very different values even for observations done during the same night, with very high uncertainties (often $\pm 0.4$). Such variations are not physical, with the values in the H and K bands being very stable, so we decided to fix $w_R = 0.55$.
The results are presented in Table~\ref{tab:oimodeler}.

In addition, we noticed that the estimated errors coming from \oimodeler~are incredibly small ($<1$\,$\mu$as). To compute more reasonable error estimates, we employed the $\chi^2_\textrm{r}+1$-method: after accepting the solution from \oimodeler~(with $\chi^2_\textrm{r,best}$), we varied each of the parameters separately, and as its error we took the maximum offset where $\chi^2_\textrm{r} \le \chi^2_\textrm{r,best} +1$.

\begin{table}[]
    \centering
    \caption{Limb-darkening coefficients $A$ and $B$ used for modelling.}
    \begin{tabular}{lrcrr}\hline\hline
 Filter     & $\lambda_c$\,[nm] & Johnson & \multicolumn{1}{c}{$A$} & \multicolumn{1}{c}{$B$} \\\hline
 WIRE       & 527   & V & 0.2613 & 0.3202 \\
 BRITE:red  & 620   & R & 0.2091 & 0.2791  \\
 BRITE:blue & 425   & B & 0.3153 & 0.3618  \\\hline
 MIRC--X    & 1618  & H & 0.0618 & 0.1646 \\
 MYSTIC     & 2170  & K & 0.0563 & 0.1384 \\
 SPICA      & 720   & R & 0.2091 & 0.2791 \\\hline
    \end{tabular}
    \tablefoot{For all of the filters, the closest Johnson equivalent was taken (based on the central wavelength, $\lambda_c$) to find the coefficients from \citet{Claret2011}. For WIRE data, the wavelength of the Hipparcos passband was assumed.
    }
    \label{tab:LDcoeffs}
\end{table}

\begin{table*}[]
    \centering
    \caption{Results of position fitting using \oimodeler. }
    \begin{tabular}{r Sr Sr Sr Sr Sr Sr Sr Sr} \hline\hline
\multicolumn{1}{c}{MJD} & \multicolumn{1}{c}{$\theta_1$ }& \multicolumn{1}{c}{$\theta_2$} & \multicolumn{1}{c}{$x$} & \multicolumn{1}{c}{$y$} & \multicolumn{1}{c}{$w_\textrm{{H}}$} & \multicolumn{1}{c}{$w_\textrm{{K}}$} & \multicolumn{1}{c}{$\chi^2_\textrm{{r}}$} \\ 
\multicolumn{1}{c}{[d]} & \multicolumn{1}{c}{[mas]} & \multicolumn{1}{c}{[mas]} & \multicolumn{1}{c}{[mas]} & \multicolumn{1}{c}{[mas]} & \\\hline 
60232.4493 & $0.972^{+0.011}_{-0.012}$ & $0.912^{+0.013}_{-0.013}$ & $ -2.003^{+0.011}_{-0.016}$ & $0.238^{+0.012}_{-0.011}$ & $0.526^{+0.003}_{-0.003}$ & $0.526^{+0.003}_{-0.003}$ & $6.747$ \\
60232.5176 & $1.046^{+0.008}_{-0.008}$ & $1.002^{+0.009}_{-0.008}$ & $ -2.236^{+0.010}_{-0.008}$ & $0.403^{+0.010}_{-0.012}$ & $0.525^{+0.003}_{-0.003}$ & $0.526^{+0.005}_{-0.003}$ & $7.980$ \\
60234.5388 & $1.032^{+0.018}_{-0.023}$ & $0.983^{+0.021}_{-0.025}$ & $2.377^{+0.022}_{-0.012}$ & $ -0.496^{+0.016}_{-0.021}$ & $ 0.524^{+0.020}_{-0.012}$ & $0.525^{+0.007}_{-0.007}$ & $2.639$ \\
60259.3951 & $1.018^{+0.011}_{-0.011}$ & $0.967^{+0.012}_{-0.012}$ & $1.507^{+0.013}_{-0.012}$ & $ -1.401^{+0.014}_{-0.015}$ & $ 0.526^{+0.005}_{-0.005}$ & $0.526^{+0.003}_{-0.003}$ & $2.904$ \\
60261.3824 & $1.020^{+0.003}_{-0.003}$ & $0.966^{+0.004}_{-0.004}$ & $ -1.474^{+0.007}_{-0.007}$ & $1.390^{+0.006}_{-0.008}$ & $ 0.527^{+0.001}_{-0.001}$ && $7.042$ \\
60597.4941 & $1.048^{+0.010}_{-0.010}$ & $0.995^{+0.012}_{-0.011}$ & $ -2.940^{+0.013}_{-0.013}$ & $1.526^{+0.032}_{-0.025}$ & $ 0.533^{+0.012}_{-0.014}$ & $0.525^{+0.005}_{-0.005}$ & $5.773$ \\
60614.4255 & $1.020^{+0.009}_{-0.009}$ & $0.968^{+0.009}_{-0.009}$ & $1.318^{+0.018}_{-0.017}$ & $0.185^{+0.014}_{-0.014}$ & $ 0.528^{+0.014}_{-0.012}$ & $0.527^{+0.001}_{-0.001}$ & $1.346$ \\
60614.4659 & $0.998^{+0.006}_{-0.007}$ & $0.948^{+0.007}_{-0.007}$ & $1.490^{+0.008}_{-0.008}$ & $0.088^{+0.007}_{-0.006}$ & $ 0.525^{+0.003}_{-0.003}$ & $0.526^{+0.001}_{-0.001}$ & $5.686$ \\

60976.4461 & $1.033^{+0.016}_{-0.018}$ & $0.986^{+0.018}_{-0.019}$ & $0.208^{+0.020}_{-0.020}$ & $ -0.946^{+0.009}_{-0.010}$ & $0.525^{+0.014}_{-0.009}$ & $0.525^{+0.005}_{-0.005}$ & $2.087$ \\
\hline
    \end{tabular}
    \tablefoot{The errors were estimated using a $\chi^2_\textrm{r}+1$-method. For one of the nights (MJD = 60976), SPICA was not used, whereas for MJD = 60261, the data from MYSTIC got corrupted.}
    \label{tab:oimodeler}
\end{table*}

\subsection{orbitize!}

The position of the secondary component on the sky with respect to the primary can be expressed tied to the orbital elements as
\begin{equation}
    \begin{pmatrix}
        \Delta \textrm{RA} \\ \Delta \delta
    \end{pmatrix}
    =
    \alpha \frac{1-e^2}{1+e \cos \varv}
    \begin{pmatrix}
        B & G \\
        A & F 
    \end{pmatrix}
    \begin{pmatrix}
        \cos \varv \\ \sin \varv
    \end{pmatrix},
\end{equation}
where $\alpha = a/d$ is the angular semi-major axis. Defining $\Omega$ as the position angle of ascending node, i.e. orientation of the ellipse on the sky, $A$, $B$, $F$, and $G$ represent the Thielle-Innes constants (denoting the $\sin$ as $s$ and $\cos$ as $c$ of the corresponding element), 
\begin{equation}
    \begin{pmatrix}
        B & G \\
        A & F 
    \end{pmatrix}
    =
\begin{pmatrix}
      c_\omega s_\Omega + s_\omega c_\Omega c_i &
     -s_\omega s_\Omega + c_\omega c_\Omega c_i \\
      c_\omega c_\Omega - s_\omega s_\Omega c_i &
     -s_\omega c_\Omega - c_\omega s_\Omega c_i
\end{pmatrix}.
\end{equation}
To derive orbital parameters from positions, we utilised the orbit-fitting software \orbitize~ \citep{orbitize}. We optimised for the following parameters: the semi-major axis, $a$, eccentricity, $e$, inclination, $i$, argument of periastron, $\omega$, parallax, $\pi$, position angle of the ascending node, $\Omega$, total mass, $M_T$, and epoch of periastron passage,~$\tau$. In total, we used 60 walkers to compute 500\,000 orbits and sampled the posterior on the last 100\,000 of them. 

\begin{figure}
    \centering
    \includegraphics[width=0.95\linewidth]{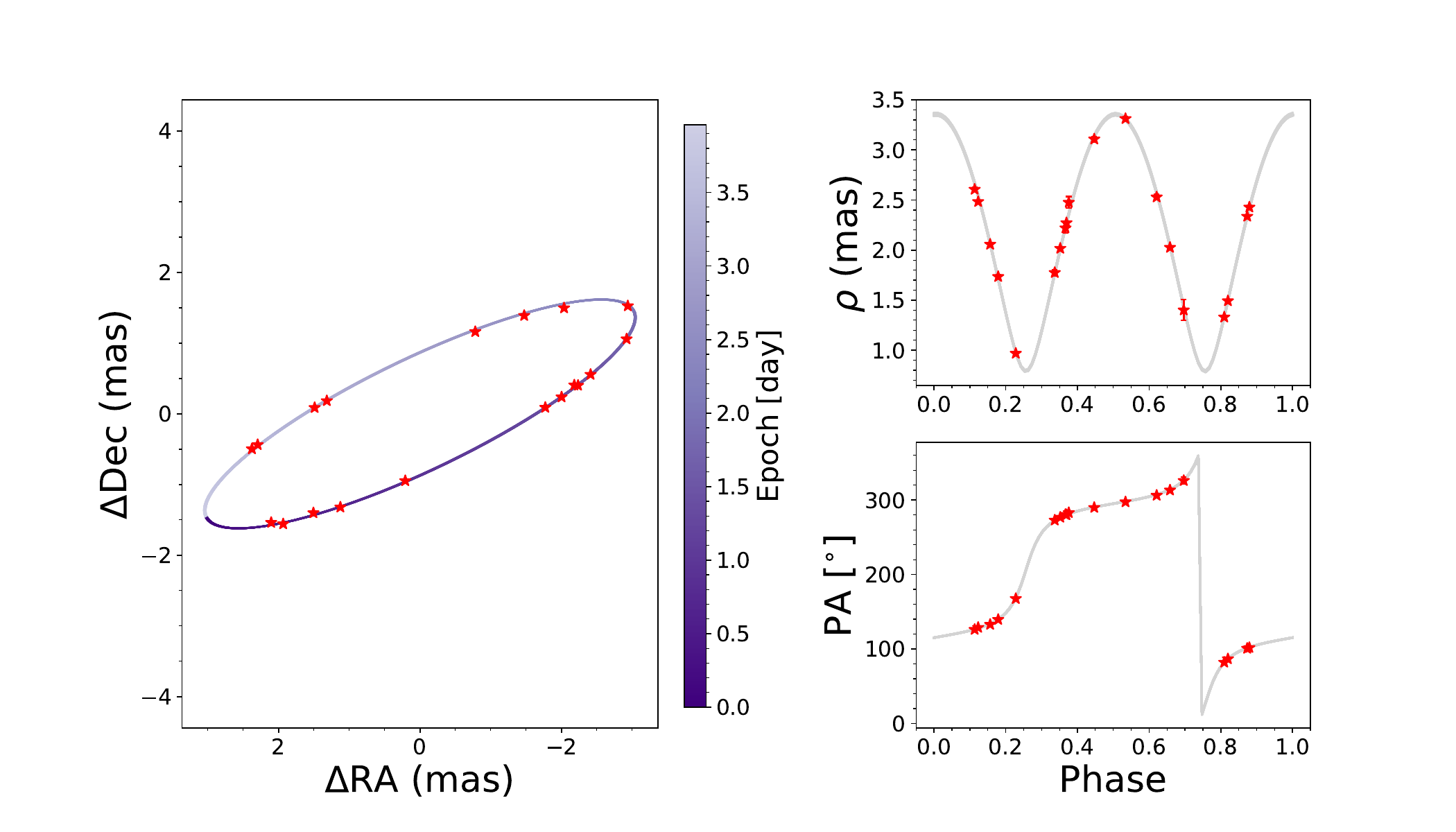}
    \caption{Orbital solution based on derived positions using \orbitize, displaying 100 orbits sampled from the result posterior compared to the data. The orbit is shown as projected on the sky (bottom left) along with the mutual distance, $\rho$, and position angle, $PA$, as a function of the orbital phase (bottom right).}
    \label{fig:orbitize}
\end{figure}

A sample of orbits from the posterior is shown in Fig. \ref{fig:orbitize}, and the result is summarised in Table \ref{tab:orbitize}. The parametrisation used by \orbitize~is optimised for observations of resolved exoplanets, usually on long-period orbits. The semi-major axis, parallax, and total mass are highly correlated. Furthermore, due to the small, but non-zero eccentricity, the position of the periastron (defined by $\tau$ and $\omega$) is only weakly constrained, resulting in a heavy correlation between these two parameters.  

Lastly, it is worth pointing out that deriving an orbit from positions alone also gives mirror solutions -- solutions with $\omega' = \omega +180 \degr$ and $\Omega' = \Omega + 180 \degr$ are also valid. To determine which of these is the true one, it is necessary to include more information coming from RVs and light curves.

\begin{table}[h]
    \centering
    \caption{Astrometric solution, as obtained from \orbitize}.
    \begin{tabular}{l|rrr}\hline\hline
    Parameter    &  \multicolumn{1}{c}{Value} &  \multicolumn{1}{c}{$\sigma_-$} & \multicolumn{1}{c}{$\sigma_+$}\\\hline
$a$\,[\rsun] & 17.68 & 0.05 & 0.06\\
$e$   & 0.0025     & 0.0017 & 0.0016 \\
$i$\,[\degr]  & 76.7318 & 0.0012 & 0.0012\\
$\omega$\,[\degr]  & 92.8  & 0.3 & 2.9 \\
$\Omega$\,[\degr]  & 115.1161 &  0.0012 & 0.0012\\
$\tau$& 0.75 & 0.34 & 0.04\\
$\varpi$\,[mas]  &  40.82 & 0.16 & 0.14\\
$M_\textrm{tot}$\,[\msun]  & 4.72 & 0.04 &  0.05\\ \hline
    \end{tabular}
    \label{tab:orbitize} 
\end{table}

\section{Combined solution}
\label{sec:robust}
In order to precisely determine the parameters of \betaur, we developed a model combining the light curve and RV modelling software \texttt{ellc} \citep{ellc} with \oimodeler, permitting us to compute a single $\chi^2$ metric and perform an MCMC posterior sampling.

Within our model, we imposed a Keplerian binary orbit with fixed parameters and allowed the following 12 parameters to be free:
\begin{itemize}
    \item individual masses, $M_1$, $M_2$,
    \item individual radii, $R_1$, $R_2$,
    \item temperature ratio, $T_2/T_1$,
    \item period, $P$,
    \item passage of periastron, $T_\textrm{peri}$,
    \item orbital inclination, $i$,
    \item eccentricity terms, $f_c = \sqrt{e} \cos\omega$ and $f_s = \sqrt{e} \sin\omega$,
    \item angular semi-major axis, $\alpha$, and
    \item position angle of the ascending node, $\Omega$.
\end{itemize}

Naturally, a different parametrisation is possible. We tried to balance the physical properties of the stars (fundamental parameters $M_n$, $R_n$), their convergence ($f_c$ and $f_s$ should, for this small eccentricity, converge better than $e$ and $\omega$), and which observables they constrain ($\alpha$ and $\Omega$ are only relevant to interferometric data, whereas another parameter such as the distance would affect the metric of all observables).

Within the code, internal parameters are derived, such as the semi-major axis,
$a = \sqrt[3]{\frac{G(M_1+M_2)P^2}{4 \pi^2}} $,
angular diameters,
$\theta_n = 2 R_n \frac{\alpha}{a}$, and
time of minimum,
$T_\textrm{min} = T_\textrm{peri} + \frac{P}{2 \pi} f(\omega, e, i)$. 
Here, the factor $f(\omega, e, i)$ comes from minimising the visual separation of the two stars.
Furthermore, the flux ratio for each passband is computed from the central wavelength (as listed in Table \ref{tab:LDcoeffs}) using Planck's approximation:
\begin{equation}
    f_\lambda = \frac{\exp\left[ \frac{ch}{\lambda k_B T_1} \right] - 1}
                {\exp\left[ \frac{ch}{\lambda k_B T_2} \right] - 1}
                \left(\frac{R_2}{R_1}\right)^2.
\end{equation}

\subsection{Constructing the $\chi^2$-metric}
To construct the $\chi^2$-metric, we denoted $x_j$ as the observed values (radial velocity, $RV$, magnitude, $m$, squared visibility, $V^2$, and closure phase, $\phi$), with the measured error, $\sigma_j$, and $x(t_j)$ as the corresponding model value at time $t_j$ (additionally $LC(t_j)$ normalised flux coming from \texttt{ellc}). In addition, we included two dataset-dependent additive factors, $\gamma_\textrm{d}$ and $\Delta_\textrm{d}$, adjusting the respective zero points of individual datasets. The factor $\gamma_\textrm{d}$ represents possible systematic deviations in determination of RVs, coming from different instruments. Secondly, as the binary exhibits small (but not negligible) ellipsoidal variations outside of the eclipse, the reference magnitude outside the eclipse is not well defined, and thus we introduced a floating zero point for each light curve, $\Delta_\textrm{d}$. This also alleviates possible systematics in absolute calibrations of the different filters, as the precise response function of the WIRE passband is not known \citep{Southworth2007}. As a consequence, however, it removes the absolute scale of photometry, and heavily correlates the temperature of the two components. For this reason, we fixed the effective temperature of the primary as 9350\,K \citep{Southworth2007} and performed the sampling on the ratio of temperatures $T_2/T_1$.

Instead of introducing these factors as individual parameters in the model, we opted to profile them analytically inside the computation. They were calculated at each step of the computation, taking the value that minimises the corresponding $\chi^2$, 
\begin{equation}
    \begin{split}
    \gamma_\textrm{d} &= \frac{1}{\sum_j \frac{1}{\sigma_j^{2}}} \sum_j \frac{RV_j - RV(t_j)}{\sigma_j^{2}}, \\
    \label{eq:pblum}
    \Delta_\textrm{d} &= \frac{1}{\sum_j \frac{1}{\sigma_j^{2}}} \sum_j \frac{m_j + 2.5 \log_{10} LC(t_j)}{\sigma_j^{2}}.
    \end{split}
\end{equation}

In the end, we can write the individual contributions to the $\chi^2$ metric for each of the observables as
\begin{equation}
\begin{split}    
    \chi_\textrm{RV}^2 &= \sum_{\textrm{d}}\sum_{j}^{N_\textrm{d}} \left(\frac{RV_j - RV(t_j) - \gamma_{\textrm{d}}}{\sigma_j}\right)^2 \\
    \chi_\textrm{LC}^2 &= \sum_{\textrm{d}}\sum_{j}^{N_\textrm{d}} \left(\frac{m_j + 2.5 \log_{10} LC(t_j) - \Delta_\textrm{d}}{\sigma_j}\right)^2\\
    \chi_\textrm{V2}^2 &= \sum_{\textrm{d}}\sum_{j}^{N_\textrm{d}} \left(\frac{V^2_j - V^2(t_j,\lambda_d)}{\sigma_j}\right)^2\\
    \chi_\textrm{CLO}^2 &= \sum_{\textrm{d}}\sum_{j}^{N_\textrm{d}} \left(\frac{1}{\sigma_j} \arctan \left[
    \frac{\sin(\phi_j - \phi(t_j,\lambda_d))}{\cos(\phi_j - \phi(t_j,\lambda_d))}
    \right]\right)^2 , \\
\end{split}
\label{eq:chi2}
\end{equation}
where the outer sum goes over different datasets, d, and the inner sum goes over the individual data points. Here, by dataset we mean a single RV curve, light curve, or an interferometric observation (V2 and CLO). In the last term, it is assumed that the units of the angles are handled properly. 

\subsection{Profile likelihood}
Initial calculations showed that the standard form of the log-likelihood function ($\ln \mathcal{L} = - 0.5 \sum_d \chi^2_d$) is strongly dominated by the interferometric datasets due to their large number of measurements and differing variance scales. Additionally, as shown in Table \ref{tab:oimodeler}, the individual interferometric fits of positions are, except for one instance, significantly larger than unity, indicating that the noise levels are underestimated and/or that additional systematics are present. Furthermore, the global analysis is based on four fundamentally different quantities, each with their own intrinsic level of uncertainty.

Rather than introducing predefined, fixed arbitrary weights or additional free parameters, we treated the overall variance level of each observable as a nuisance parameter. Specifically, for $k \in \{\mathrm{RV, LC, V2, CLO} \}$, we introduced a scaling factor, $w_k$. Instead of being sampled, these nuisance parameters were eliminated analytically using a profile likelihood approach, i.e. by maximising the likelihood with respect to the respective $w_k$ and eliminated by profiling. We provide a derivation of this equation in Appendix \ref{app:profile_likelihood}. This results in a likelihood of the form
\begin{equation}
\ln \mathcal{L} = - \frac{1}{2} \sum_k N_k \ln \chi^2_k,
\label{eq:profile_likelihood}
\end{equation}
with $N_k$ the number of data points in each.  The resulting form is equivalent to marginalising over an unknown variance scale using a Jeffreys prior \citep{Jeffreys1946}, $p(\sigma) \propto 1/\sigma$, and corresponds to the standard treatment of scale parameters in likelihood-based inference \citep{Sellentin2016}. This approach naturally accounts for the differing and partially uncertain noise properties of the heterogeneous observables without introducing ad hoc weighting factors.

\subsection{Establishing RV uncertainty level}
\label{sec:RV_uncertainty_level}
In this work, we combined heterogeneous RV datasets spanning more than a century altogether, each published with its own internal uncertainties, depending on the methodology of extraction. An initial model computation, which made use of the published uncertainties, revealed substantial inconsistencies between the individual datasets, particularly in the RV residuals. We therefore performed a diagnostic analysis of the residual structure and revised the RV noise model by introducing additional variance terms for the affected datasets. We would like to stress that the goal of this procedure was not to artificially reduce the total $\chi^2$ but instead to construct a robust and consistent noise model that prevents a single dataset (or even a single value) from disproportionately dominating the convergence.

An inspection of the data published by \citet{Strassmeier2020} revealed that the presented uncertainties of the individual data vary significantly and are strongly asymmetric between the two components. While the velocities of the primary were reported with a precision often better than $0.4$\,\kms~(in some cases even 0.06\,\kms), for the secondary the precision was consistently larger than $0.6$\,\kms, and in some cases exceeded 5\,\kms. We note that this asymmetry is not present in other datasets (e.g. \citet{Behr2011}, or the NARVAL datasets), where the uncertainties between the components are very similar.

To asses the consistency of all the RV datasets, we computed a diagnostic fit using only RVs, letting only the masses of the two components and the epoch vary, while fixing all remaining parameters to the values of \citet{Southworth2007}. We evaluated the individual contribution to the $\chi^2$ for each dataset and component. As shown in Table \ref{tab:RVnois}, the contributions differ drastically, indicating that some of the reported uncertainties are likely underestimated. In particular, the primary RVs of the dataset D completely dominate the $\chi^2$.

To ascertain an internal consistency within the RV dataset, we introduced an additional variance term ('jitter'), $\varsigma$, quadratically added to the nominal uncertainties. For each dataset, we estimated the level of jitter required to bring $\chi^2/N = 1$ of the particular dataset within this diagnostic fit. We used this procedure only to identify which datasets contain significantly underestimated uncertainties.

Based on this analysis, we applied a fixed individual jitter term to the two most discrepant datasets (B and D), $\varsigma_\textrm{B} =$ 2\,\kms and $\varsigma_\textrm{D} =$ 0.8\,\kms, respectively, and introduced a global jitter term, $\varsigma_\textrm{global} = $ 0.4\,\kms, to account for the variance not captured by the nominal uncertainties. The total uncertainty used within the model is thus
\begin{equation}
    \sigma_\textrm{RV,d,j} = \sqrt{\sigma_\textrm{nom,j}^2 + \varsigma_\textrm{global}^2+\varsigma_\textrm{d}^2}.
\end{equation}
This approach ensured that unrealistically small reported uncertainties did not dominate the likelihood, while preserving the relative weighting of measurements within each dataset. Furthermore, these terms were determined before further MCMC convergence and were not changed afterwards, to avoid introducing additional free parameters or iterative tuning.

\begin{table}[]
    \centering
    \caption{Uncertainty level on individual RV datasets.}
    \begin{tabular}{l|lcr|lllc}\hline\hline
Ds	&	Cmp	& $N$& $\chi^2/N$& $\sigma_\textrm{min}$& $\sigma_\textrm{avg}$& $\sigma_\textrm{max}$& $\varsigma$\\	
& & & & \multicolumn{4}{c}{[\kms]}\\\hline
A	&	both	&	92	&	0.97	&	5	&	5	&	5	&	0	\\	
	&	prim	&	46	&	0.89	&	5	&	5	&	5	&	0	\\	
	&	sec	&	46	&	1.05	&	5	&	5	&	5	&	1.131	\\	\hline
B	&	both	&	40	&	9.04	&	0.7	&	0.7	&	0.7	&	1.984	\\	
	&	prim	&	20	&	12.43	&	0.7	&	0.7	&	0.7	&	2.366	\\	
	&	sec	&	20	&	5.29	&	0.7	&	0.7	&	0.7	&	1.450	\\	\hline
C	&	both	&	40	&	6.01	&	0.149	&	0.26	&	0.556	&	0.492	\\	
	&	prim	&	20	&	7.74	&	0.151	&	0.262	&	0.550	&	0.551	\\	
	&	sec	&	20	&	4.28	&	0.149	&	0.257	&	0.556	&	0.423	\\	\hline
D	&	both	&	246	&	10.14	&	0.056	&	1.081	&	6.931	&	0.803	\\	
	&	prim	&	123	&	18.56	&	0.056	&	0.424	&	3.949	&	0.704	\\	
	&	sec	&	123	&	1.73	&	0.629	&	1.739	&	6.931	&	1.081	\\	\hline
E	&	both	&	36	&	8.83	&	0.116	&	0.172	&	0.330	&	0.415	\\	
	&	prim	&	18	&	7.12	&	0.116	&	0.167	&	0.282	&	0.343	\\	
	&	sec	&	18	&	10.54	&	0.122	&	0.177	&	0.330	&	0.478	\\	\hline
    \end{tabular}
    \tablefoot{The following values are listed, split per dataset, Ds, and component, Cmp.: the number of pieces of data, $N$, the contribution per data point to the metric $\chi^2/N$, the minimum, average, and maximum nominal uncertainties, $\sigma$, and the jitter, quadratically added to the individual uncertainties so that the total $\chi^2/N = 1$. The datasets are as follows: A -- \citet{Baker1910}, B -- \citet{Smith1948}, C -- \citet{Behr2011}, D -- \citet{Strassmeier2020}, and E -- NARVAL spectra.
    }
    \label{tab:RVnois}
\end{table}

\subsection{Light curve adjustments}
To account for limb darkening in our model, we compared the central wavelength of each passband (or, in the case of the WIRE data, to the Hipparcos $H_p$) with the closest Johnson equivalent and adopted the corresponding values of the coefficients from \citet{Claret2011}, listed in Table \ref{tab:LDcoeffs}. In the case of the high-precision WIRE data, the model based solely on limb darkening was not properly reproducing the overall morphology of the eclipses, leaving systematic residuals located symmetrically at the edges of the eclipses. This suggested that additional surface-brightness effects, neglected in the simplified model, are significant at the precision of this particular dataset.

To improve the fit of the WIRE photometry, we included gravity darkening using the coefficient $y = 0.5544$ from \citet{Claret2011}. We additionally explored the sensitivity of the residual structure to the reflection coefficient and adopted a value of 0.2, which minimised the systematic residuals at the edges of the eclipses. While this treatment remains approximate, the resulting corrections mainly affect the detailed eclipse morphology rather than the system geometry. These corrections were applied only to the WIRE photometry, whose substantially higher precision made the residual structure clearly detectable, while their effect on the BRITE  light curves with lower precision was negligible within the observational uncertainties.

The computation time was heavily dominated by the WIRE dataset. To reduce it, we took into account only every fiftieth point lying outside of eclipse while keeping eclipse phases fully sampled. This reduced the number of points from 30015 to 4804 and the computation time by a factor of approximately 3 to 4.

\subsection{$\chi^2$ maps}
To ensure that the initial positions of the walkers are located near the global minimum before performing a sampling with MCMC, we performed a series of $\chi^2$ maps, where we kept the majority of the parameters fixed, and varied one or two of them. As a baseline, we assumed that the solution by \citet{Southworth2007} is close in the parameters not tied to interferometry -- all except $\alpha$ and $\Omega$. These, in turn, could be taken from the astrometric orbit performed with \orbitize.

\begin{figure}
    \centering
    \includegraphics[width=.9\linewidth]{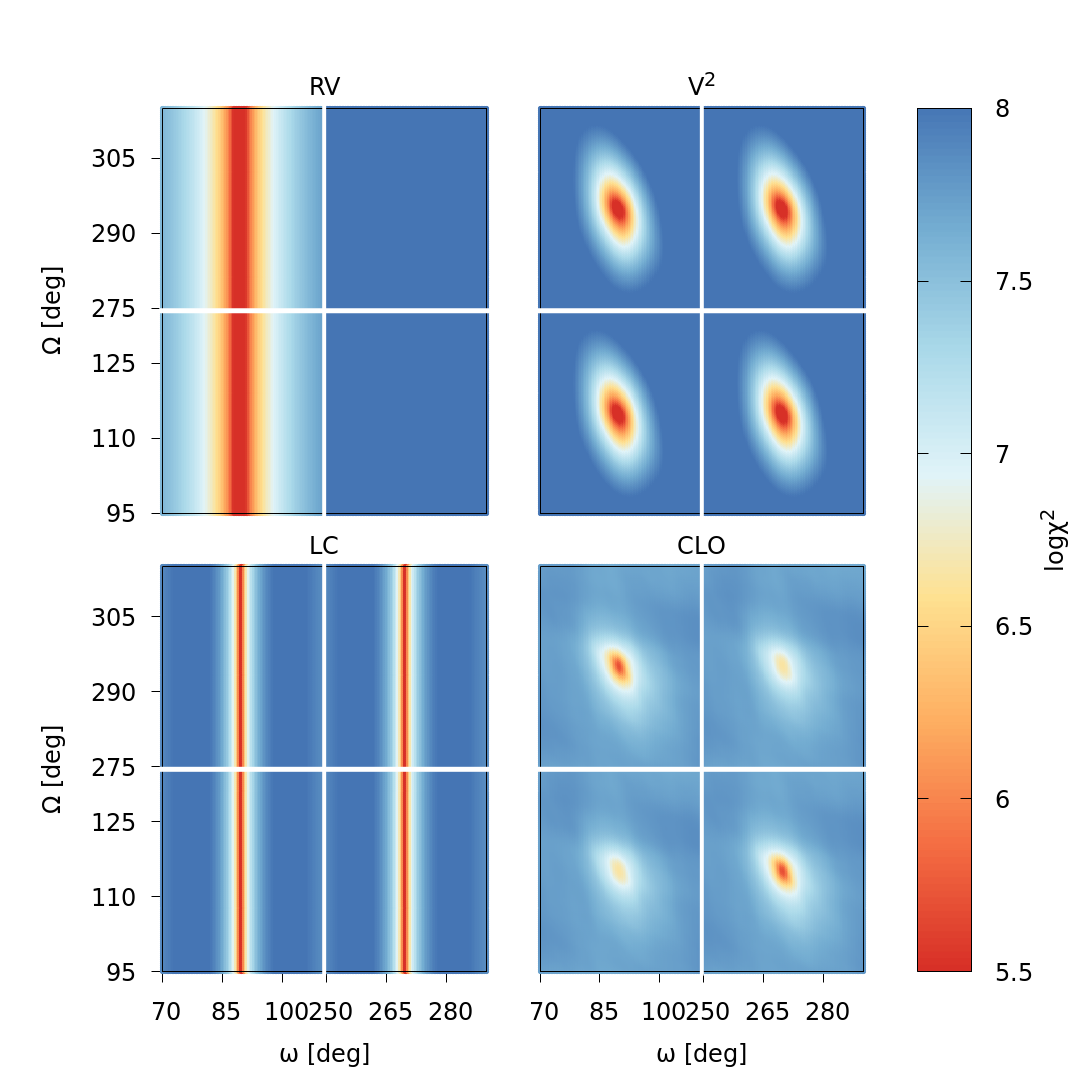}
    \caption{$\chi^2$--map with varying values of argument of periastron, $\omega$, and positional angle, $\Omega$, in the vicinity of the possible mirror solutions and step~0.4\degr. The values of $\chi^2$ were calculated separately for the four different observables -- RV, light curves (LCs), squared visibilities, V$^2$, and closure phases (CLOs), and are colour-coded in log-scale, with a red colour indicating lower values of the metric. For graphical reasons, the values for RV-dataset were multiplied by a factor of 100.}
    \label{fig:chi2map}
\end{figure}

In particular, we verified whether the solution from \orbitize~agrees with the model by performing a $\chi^2$--map between the argument of periastron, $\omega$, and position angle, $\Omega$. Focusing on the possible mirror solutions $\omega \in \{ 90\degr, 270\degr\}$ and $\Omega \in \{115\degr, 295\degr\}$, we mapped the corresponding neighbourhoods of $\pm 20\degr$ with step-size of $0.2\degr$. We demonstrate this in Fig.~\ref{fig:chi2map}, where we show that, as expected, the value of $\Omega$ does not affect the value of  $\chi_\textrm{RV}^2$ and $\chi_\textrm{LC}^2$. Furthermore, while the light curve does not distinguish between $\omega$ , the RVs strongly prefer $\omega \approx 90\degr$. Similarly, $V^2$ does not show a clear preference between these four solutions, whereas the closure phases show much deeper minima for $(\omega, \Omega) \approx (90\degr, 295\degr)$ and $\approx (270\degr, 115\degr)$ than for the remaining two solutions. Based on this analysis, in further computations we considered the solution $(\omega, \Omega) \approx (90\degr, 295\degr)$. 

\subsection{MCMC sampling}

We performed MCMC sampling, using 80 walkers and 6\,000 steps, out of which we sampled the posterior on the last 1\,000. The result is presented in Table \ref{tab:result}, along with the allowed range for each of the parameters, the converged value, and the respective errors. For reference, we also list four parameters not directly minimised in our solutions -- the semi-major axis, $a$, eccentricity, $e$, argument of periastron, $\omega$, and distance, $d$. Their respective errors were estimated through uncertainty propagation, with the larger value of $(\sigma_-,\sigma_+)$ used. 

\begin{figure}
    \centering
    \includegraphics[width=0.85\linewidth]{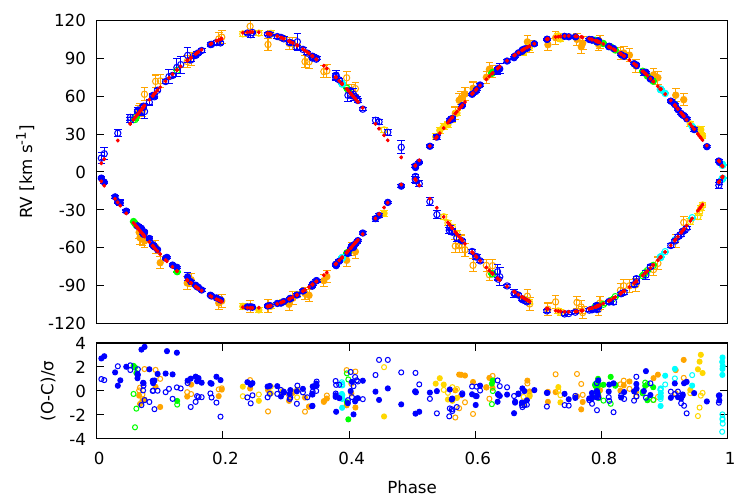}
    \caption{Comparison between the observed and synthetic (red) RVs, along with the residuals $(O-C)/\sigma$, plotted with respect to orbital phase. Colours represent the different datasets -- orange: \citet{Baker1910}, yellow: \citet{Smith1948}, green: \citet{Behr2011}, blue: \citet{Strassmeier2020}, and cyan: NARVAL measurements.}
    \label{fig:RVs}
\end{figure}

\begin{figure}
    \centering
    \includegraphics[width=0.85\linewidth]{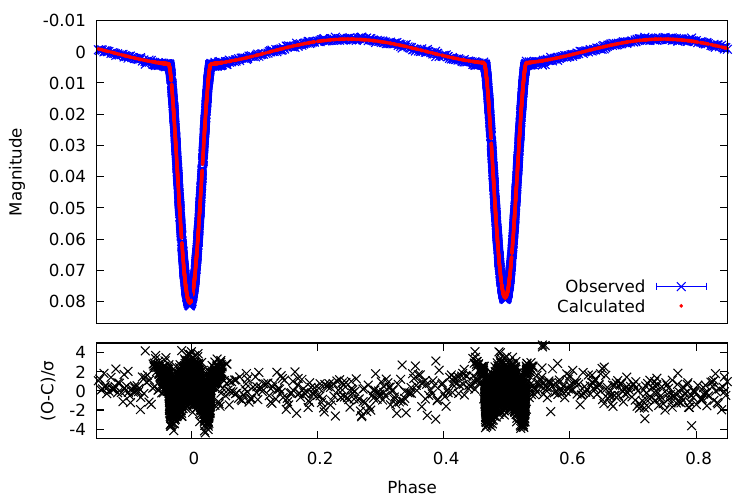}\\
    \includegraphics[width=0.85\linewidth]{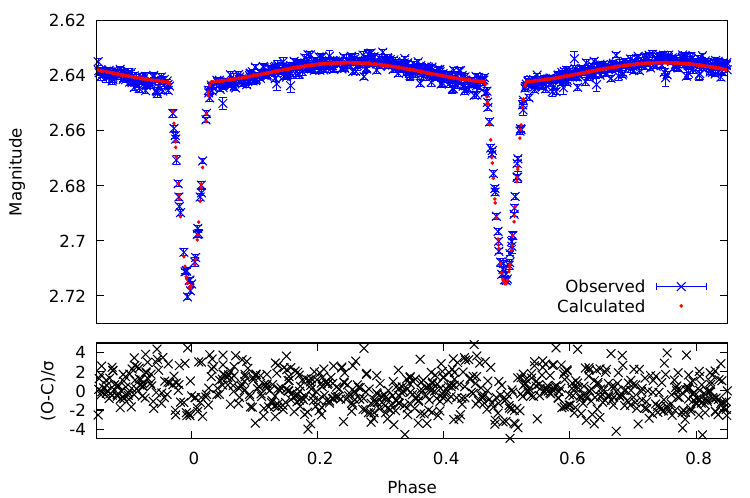}\\
    \includegraphics[width=0.85\linewidth]{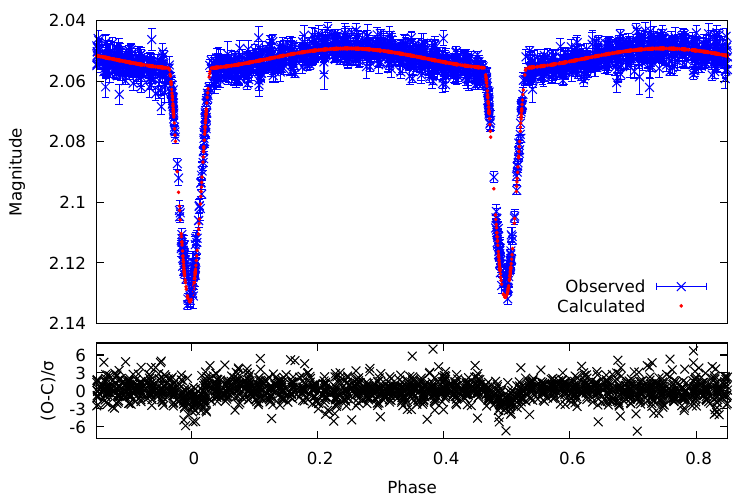}
    \caption{Comparison between the observed (blue) and synthetic light curves calculated by our model (red), along with the values of residuals (black), all plotted with respect to the orbital phase: (top) WIRE, (centre) BRITE:blue, and (bottom) BRITE:red.}
    \label{fig:LCs}
\end{figure}

\begin{figure}
    \centering
    \includegraphics[width=0.85\linewidth]{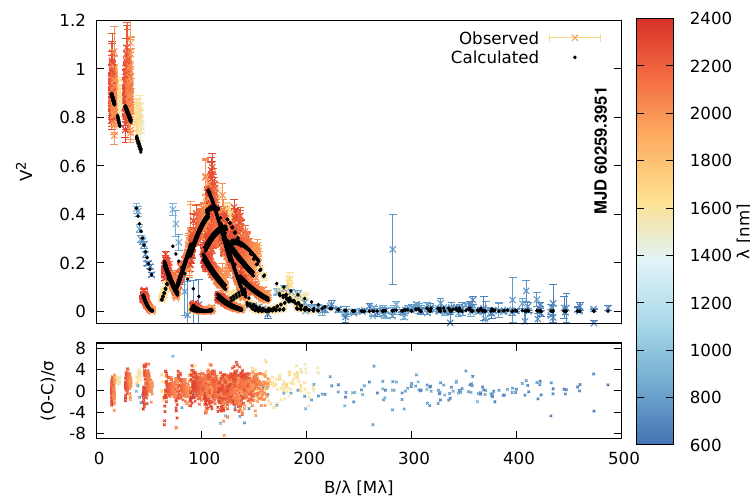}\\
    \includegraphics[width=0.85\linewidth]{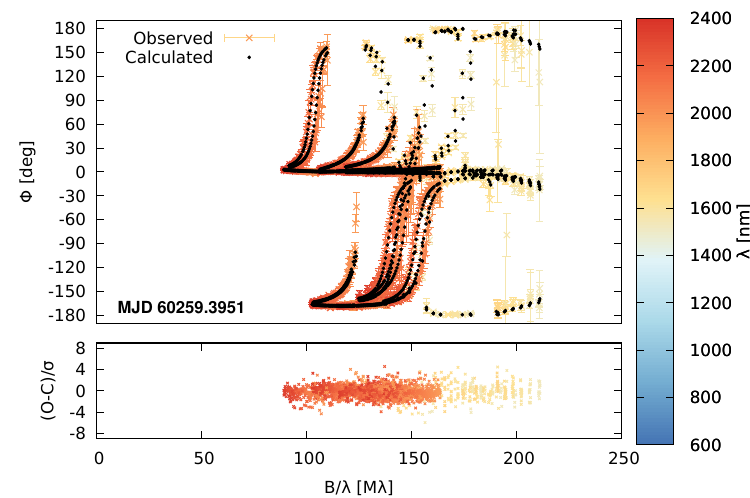}
    \caption{Comparison between the observed and calculated (black) squared visibilities (top) and closure phases (bottom) corresponding to the observation of November 11 2023, along with the respective residuals. The data are colour-coded based on wavelength.}
    \label{fig:IFs}
\end{figure}

The comparison between the data (RVs, LCs, $V^2$s, and closure phases) and our model can be seen in Figures \ref{fig:RVs}, \ref{fig:LCs}, and \ref{fig:IFs}, respectively, along with the individual residuals $(O-C)/\sigma$. Furthermore, Fig. \ref{fig:solution} shows the resulting corner plot of our solution. The individual contributions from each dataset to the $\chi^2$-metric are listed in Table \ref{tab:result_chi2}. 

Nonetheless, the MCMC sampling did not produce a multimodal solution, and all walkers eventually converged. In Fig. \ref{fig:trace}, we show the evolution of four parameters (the masses, $M_1$ and $M_2$, the radius of the primary component, $R_1$, and the angular semi-major axis, $\alpha$), as a function of the steps within this sampling. To verify the stability of the solution, we split the last 3\,000 steps into chunks of one thousand steps and extracted the values and the statistical deviations from each separately, as well as their combinations, and found no significant differences.

\begin{table*}[h]
\centering
\begin{minipage}{0.65\textwidth}
\centering
\caption{Converged MCMC solution.}
\begin{tabular}{ll|rr|rrr} \hline\hline
\multicolumn{2}{c|}{Param.}   & \multicolumn{1}{c}{Min} & \multicolumn{1}{c}{Max} & \multicolumn{1}{|c}{Result}        & \multicolumn{1}{c}{$\sigma_-$}  & \multicolumn{1}{c}{$\sigma_+$} \\\hline
$M_1$&	\msun	&	2.2	&	2.5	&	2.3589	&	0.0047	&	0.0044	\\	
$M_2$&	\msun	&	2.1	&	2.4	&	2.2927	&	0.0043	&	0.0042	\\	
$R_1$&	\rsun	&	2.5	&	3.1	&	2.7515	&	0.0017	&	0.0016	\\	
$R_2$&	\rsun	&	2.3	&	2.8	&	2.6220	&	0.0017	&	0.0016	\\	
$P$&	d	&	3.9600	&	3.9601	&	3.960046633	&	2.30E-08	&	2.51E-08	\\	
$T_\textrm{peri}$&	d	&	53827.1	&	53827.3	&	53827.2048	&	0.0030	&	0.0026	\\	
$T_2/T_1$&		&	0.90	&	1.02	&	0.997543	&	0.000087	&	0.000096	\\	
$i$&	\degr	&	76	&	78	&	76.7875	&	0.0020	&	0.0020	\\	
$\alpha$&	mas	&	3.1	&	3.5	&	3.36514	&	0.00023	&	0.00024	\\	
$f_c$&		&	-0.01	&	0.01	&	-0.00058	&	0.00017	&	0.00019	\\	
$f_s$&		&	0.03	&	0.06	&	0.04025	&	0.00044	&	0.00045	\\	
$\Omega$&	\degr	&	290.0	&	305.0	&	295.1466	&	0.0056	&	0.0051	\\	\hline
$a$&	\rsun	&		&		&	17.583	&	0.011	\\			
$e$&		&		&		&	0.00162	&	0.00004	\\			
$\omega$&	\degr	&		&		&	90.83	&	0.24	\\			
$d$&	pc	&		&		&	24.299	&	0.015	\\			\hline
\end{tabular}
\tablefoot{The table lists the fitted parameters, their allowed range, and the errors, $\sigma_-$ and $\sigma_+$.}
\label{tab:result}
\end{minipage}
\end{table*}

\begin{table}[h]
    \centering
    \caption{Contribution of each dataset to the $\chi^2$ of the final solution.}
    \begin{tabular}{l|rrrrr}\hline\hline
        Obs. & Ds & \multicolumn{1}{c}{$N$}   & \multicolumn{1}{c}{$\chi^2$} & \multicolumn{1}{c}{$\chi^2/N$}& \multicolumn{1}{c}{$\chi^2_\textrm{r}$} \\\hline
RV	&	A	&	92	&	85	&	0.918	&		\\
	&	B	&	40	&	44	&	1.088	&		\\
	&	C	&	40	&	45	&	1.116	&		\\
    &	D	&	246	&	260	&	1.058	&		\\
	&	E	&	36	&	73	&	2.026	&		\\
    \cline{2-6}
	&	All	&	454	&	506	&\	1.114	&\	1.144	\\\hline
LC	&	L	&	4804	&	8729	&	1.817	&		\\
	&	M	&	708	&	2213	&	3.125	&		\\
	&	N	&	1824	&	4717	&	2.586	&		\\
    \cline{2-6}
	&	All	&	7336	&	15659	&	2.135	&	2.138	\\\hline
V2	&	R	&	873	&	9096	&	10.419	&		\\
	&	H	&	3851	&	22408	&	5.819	&		\\
	&	K	&	7319	&	28775	&	3.932	&		\\
    \cline{2-6}
	&	All	&	12043	&	60279	&	5.005	&	5.010	\\\hline
CLO	&	H	&	2921	&	6914	&	2.367	&		\\
	&	K	&	9009	&	15323	&	1.701	&		\\
    \cline{2-6}
	&	All	&	11930	&	22237	&	1.864	&	1.866	\\\hline
Total	&		&	31763	&	98680	&		&	3.108	\\\hline
\end{tabular}
    \tablefoot{For each dataset the number of points, $N$, total $\chi^2$ contribution, and the reduced value, $\chi^2_\textrm{r}$, for the entire observable are listed. The datasets are as follows: A -- \citet{Baker1910}, B -- \citet{Smith1948}, C -- \citet{Behr2011}, D -- \citet{Strassmeier2020} and E -- NARVAL spectra; L -- WIRE data \citep{Southworth2007}, M -- BRITE red and N -- BRITE blue (both \citet{Strassmeier2020}); R -- SPICA, H -- MIRC/MIRC--X, and K -- MYSTIC.}
    \label{tab:result_chi2}
\end{table}

\section{Discussion}
\subsection{Model approximations and agreement}
In this work, the main focus is the treatment of interferometric data and their inclusion in simultaneous modelling with RVs and light curves. We neglected to perform an optimisation of the effects of limb- and gravity darkening (LD and GD, respectively) on the light curves, assuming the respective table values, and considered the effect of GD only on the WIRE dataset, where its effect is not negligible. Furthermore, the values of the LD and GD coefficients are only approximations coming from the central wavelengths of the used passbands, and adopting the closest Johnson filter values. Naturally, a further step would be to include LD coefficients as free parameters in the MCMC sampling. This would, however, greatly expand the parameter space by 12 additional and weakly constrained parameters and is beyond the intended scope of this article. 

Additionally, in our model, we kept the temperature of the primary at a fixed value and only considered the temperature ratio $T_2/T_1$ as a free parameter. Separating these two parameters would cause a heavy correlation between them. In fact, this effect is even strengthened in this model, as we considered only relative light curves (the arbitrary term $\Delta_d$ defined in Eq. \ref{eq:chi2}), making an independent measurement of $T_1$ impossible.

Although the combined model reproduces all of the observables well globally, small residual structures remain visible. They are addressed separately for the different observables.
\begin{enumerate}
\item RV. Due to the very heterogeneous datasets containing underestimated uncertainties, we constructed a robust noise model by introducing a jitter term as an additional variance in the model. The residual show a small phase-correlated structure, mainly at the phases with low velocity separation, and likely reflects remaining systematic effects associated with line blending. 
The model includes the Rossiter--McLaughlin effect \citep{Rossiter1924,McLaughlin1924}, natively implemented in \ellc, where we are assuming synchronous rotation. However, due to the grazing nature of the eclipses, this effect is overall limited on the global solution. Additional physical effects, such as a shift in the observed RV caused by the mutual irradiation and other proximity effects, are not explicitly included in the model, and may contribute to the RV residuals. The presence of such effects gives further justification for introducing a jitter term to the reported uncertainties described in Sect. \ref{sec:RV_uncertainty_level}.

\item LC. Small systematic structures can also be seen in the residual plots of the light curve datasets, most visible in the WIRE dataset. As they are, to a high degree, symmetric around the start and end of the eclipse, they likely reflect the simplified treatment of surface brightness implemented in the model, rather than an inaccuracy of the orbital solution.
\item Interferometry. The values of the reduced $\chi^2$ are significantly above unity for each of the $V^2$ datasets. Nonetheless, the global structure is well reproduced, and the result is strengthened by the much better reproduction of closure phases, which are much less sensitive to calibration uncertainties than the squared visibilities. This suggests that the discrepancies are dominated by interferometric systematics, possibly coming from transfer function instability, rather than deficiencies in the binary model. Furthermore, calibration issues can also explain the slight overshooting of the data above unity at the shortest frequencies. This effect also causes discrepancies in estimation of the angular diameter of the two stars between the individual nights, as shown in Table \ref{tab:oimodeler}. Overall, this might be partially caused by a substantial difference in brightness between the calibrators and the science target, which can affect the stability of the calibration. In addition, the interferometric observables show a systematic decrease in reduced $\chi^2$ with increasing wavelength. This could, at least to a certain degree, come from the limitations of the simplified chromatic flux-ratio treatment, where we assumed a Planck-law approximation. While introducing independent flux ratios for each passband would provide more flexibility for the model, it would substantially increase the model degeneracy. As such, we retain the physically coupled temperature-based model.
\end{enumerate}

In our model, a Keplerian orbit of two bodies with constant parameters was assumed. Though the existence of a distant tertiary cannot be formally rejected, we found no indication in the data supporting its presence. In particular, the work of \citet{Southworth2007} already detected no third light even at the mmag-precision of the WIRE data, and neither the RV residuals nor the interferometry measurements indicate an additional component. Furthermore, the timings of eclipses compiled in the O-C Gateway\footnote{\url{https://var.astro.cz/en/Stars/9278}, accessed June 1, 2026}  do not indicate significant secular effects over more than a century of observations. 
We also explored the possibility of an apsidal motion. However, due to the nearly circular orbit of \betaur, $\omega$ is already weakly defined and highly correlated to $T_\textrm{peri}$. The addition of this extra parameter produced a highly degenerate and ill-defined solution.

To incorporate the miscalibration present in the interferometric data, we adjusted the wavelength scale using Eq. (\ref{eq:mm_miscal}). In addition to the correction, as per the MIRC--X/MYSTIC manual, one must also include a systematic error of $0.2\%$ to parameters related to angular size. In our case, this would correspond to the two radii and the angular semi-major axis (and, in turn, the distance). However, in this specific case, the angular semi-major axis is the most directly affected by this miscalibration, while the radii are constrained by the high-precision light curves. We therefore added in quadrature the suggested $0.2\%$ to the nominal uncertainty of the derived distance.

In Table \ref{tab:fundamental_parameters}, we compare the physical parameters derived from our solution with the result of \citet{Southworth2007}. Notably, both the mass and radius of the primary as well as the mass of the secondary are in excellent agreement between the two solutions. Our distance is slightly lower, but still within a $1\sigma$ interval. The largest discrepancy between the two studies is in the radius of the secondary, which differs by approximately $2.5\sigma$.

The precision of the present solution is a substantial improvement on the previous result, primarily due to the combined global modelling of all observables with the inclusion of long-baseline interferometry and rich historic data spanning more than a century. At the same time, the parameter precision is consistent with modern interferometric studies on binary systems, such as \citet{Gallenne2019}, \citet{Lester2019A}, or \citet{Danner2025}.
\begin{table}[h!]
    \centering
    \caption{Comparison of stellar parameters and the distance.}
    \begin{tabular}{c|cr|cr}\hline\hline
    & \multicolumn{2}{c|}{S07} & \multicolumn{2}{c}{This work} \\\hline
 Prim. & \\\hline														
$M_1$ [\msun]	& $2.38 \pm 0.03$ & $1.1	\%$& $2.359	\pm	0.005$ & $0.20	\%$\\
$R_1$ [\rsun]	& $2.76 \pm 0.02$ & $0.6	\%$& $2.752	\pm	0.002$ & $0.06	\%$\\\hline
 Sec. & \\\hline												
$M_2$ [\msun]	& $2.29 \pm 0.03$ & $1.2	\%$& $2.293	\pm	0.004$ & $0.19	\%$\\
$R_2$ [\rsun]	& $2.57 \pm 0.02$ & $0.7	\%$& $2.622	\pm	0.002$ & $0.06	\%$\\\hline
$d$ [pc]	& $24.8\pm0.8$ & $3.2	\%$& $24.30	\pm	0.05$ & $0.21	\%$\\
\hline
\end{tabular}
    \tablefoot{We compare the masses and radii of the two components of \betaur, as well as the distance to the system, with the values obtained by \citet{Southworth2007}. For each of the parameters, we list the respective values together with their absolute and relative precision.}
    \label{tab:fundamental_parameters}
\end{table}

\subsection{Evolution track comparison}
We performed a comparison with stellar evolution tracks, obtained from the MESA Isochrones \& Stellar Tracks (MIST, \citet{MIST0,MIST1,MIST2}) service. Namely, we compared the evolution of the two individual stellar radii and the temperature ratio with tracks of different  iron abundances [Fe/H]. In order to calculate the corresponding temperature ratio, we interpolated the values provided by MIST into a common timeline for each pair of individual tracks.

\begin{figure}
    \centering
    \includegraphics[width=0.9\linewidth]{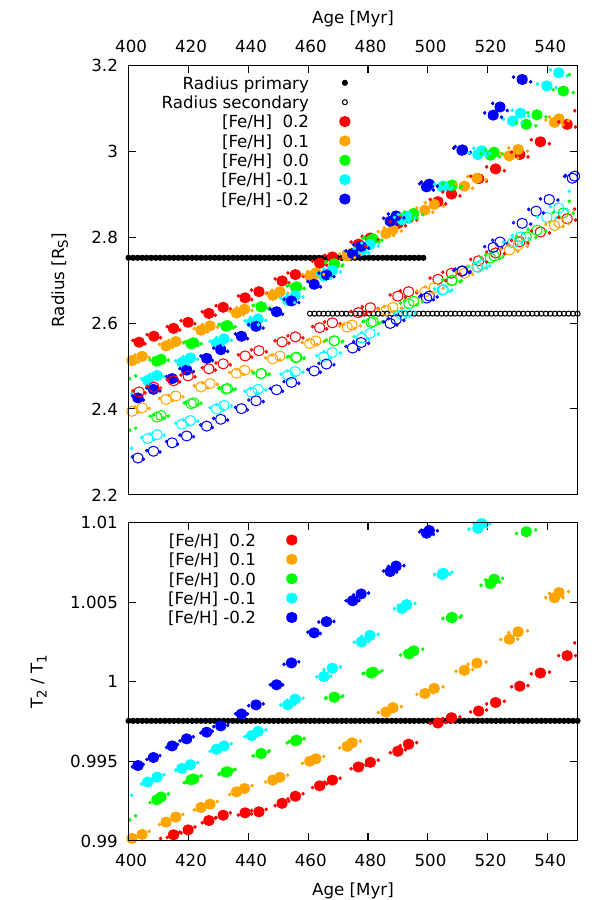}
    \caption{Comparison of our result with MIST evolution tracks: (top) the radius of the primary (filled circles) and the secondary component (empty circles); (bottom) the temperature ratio. The $1\sigma$ uncertainty in the individual masses is represented by smaller points -- evolutionary tracks of these masses. Black lines correspond to the values derived within our model. }
    \label{fig:eep}
\end{figure}

As can be seen in Fig. \ref{fig:eep}, the primary component seems to be in an evolutionary state in which the iron abundance does not affect its radius. The preferred ages based on the primary and secondary radii differ slightly, at the level of approximately 10\,Myr, but remain consistent with a coeval evolution. 
The comparison with the temperature ratio, assuming the age range based on the radii, suggests a slightly super-solar iron abundance -- [Fe/H] $\approx 0.1$.

We note that no formal fitting was performed and that this discussion is solely based on the comparison between the different tracks and our result. A proper analysis would require a more rigorous approach, finer grid of tracks, consideration of rotation effects, and comparison of different evolutionary models. This would go beyond the scope of this article. Instead, our discussion serves to demonstrate that our orbital solution provides meaningful constraints on the evolutionary state of the system, and in the future may serve as a useful benchmark for further studies, dedicated to analysis of stellar evolution and abundances; for instance, in the context of the upcoming PLATO mission \citep{PLATO}.

\section{Conclusion}
In this work, we present a detailed analysis of the eclipsing binary \betaur, for which we combined new interferometric observations with archival interferometry, RVs, and photometric data. To perform a simultaneous fitting of all observables within a single consistent solution, we developed a unified modelling framework.

We first derived an interferometric orbit based on astrometric positions obtained from interferometric modelling. This solution had already demonstrated strong correlations between key parameters and highlighted intrinsic degeneracies, such as the mirror ambiguity in the orbital orientation. These limitations were subsequently overcome by incorporating RVs and light curves into a combined model.

Our joint modelling approach, coupling interferometric modelling with the \texttt{ellc} code, allowed us to derive a coherent set of orbital and stellar parameters. The inclusion of interferometric data proved essential for resolving degeneracies that cannot be lifted by spectroscopy or photometry alone by firmly constraining the angular scale of the system. At the same time, the photometric and spectroscopic data provide strong constraints on the dynamical and physical properties of the components.

We developed a robust noise model to take into account the demonstrated underestimation of the formal uncertainties of RVs. By adopting a profile likelihood approach in the final solution, we avoided the dominance of particular datasets without adding ad hoc scaling factors, leading to realistic and statistically consistent uncertainties. 

In the present work, several simplifications were assumed, including fixed limb- and gravity-darkening coefficients and a partially constrained temperature scale. While these are justified within the scope of this study, future analyses could explore their impact by extending the parameter space or incorporating more detailed modelling of stellar atmospheres. The methodology developed here will be applied to the full sample of targets selected for the SPICA programme, enabling a homogeneous and high-precision determination of stellar parameters across a broad population of spectroscopic and eclipsing binaries.

\section*{Data availability}
The complete target list of the programme can be found at \url{https://doi.org/10.5281/zenodo.21106113}

\begin{acknowledgements}
\\
This project has received funding from the European Research Council (ERC) under the European Union’s Horizon 2020 research and innovation programme (Grant agreement No. 101019653). SPICA has been funded by CNRS, Observatoire de la Côte d'Azur, Université Côte d'Azur, Région Sud, and the University of Aarhus. 
This work is based upon observations obtained with the Georgia State University Center for High Angular Resolution Astronomy Array at Mount Wilson Observatory.  The CHARA Array is supported by the National Science Foundation under Grant No. AST-2034336 and AST-2407956. Institutional support has been provided from the GSU College of Arts and Sciences, Office of the Provost, and Office of the Vice President for Research and Economic Development.
This research has made use of the SIMBAD database  \citep{simbad} and VizieR catalogue access tool \citep{vizier}, operated at CDS, Strasbourg, France. 
This research has made use of the tools developed by Jean-Marie Mariotti Center.
We warmly thank Chris Farrington, Becky Flores, Olli Majoinen, Heven Renteria, and Norm Vargas for their excellent support during the night operations.
JJ and DM acknowledge David Mary for helpful discussion on the Jeffrey's prior. We thank Pierre Maxted for his aid with the \ellc~code. We also express our gratitude to Petr Harmanec for the initial methodology of target selection.
\\
\end{acknowledgements}
\bibliographystyle{aa}
\bibliography{ref}

\begin{appendix}
\section{NARVAL radial velocities}
\begin{table}[h]
    \caption{Radial velocities of \betaur\ from the NARVAL spectra.}
    \centering
    \begin{tabular}{c|rlrl}\hline\hline
    HJD    &  \multicolumn{2}{c}{Primary} &  \multicolumn{2}{c}{Secondary}\\
    & \multicolumn{2}{c}{RV [\kms]} & \multicolumn{2}{c}{RV [\kms]}\\\hline
$56513.6555$ & $-86.22$ & $\pm	0.14$ & $52.23$ & $\pm	0.16 $\\
$56513.6563$ & $-86.26$ & $\pm	0.13$ & $52.02$ & $\pm	0.14 $\\
$56513.6571$ & $-86.19$ & $\pm	0.14$ & $51.95$ & $\pm	0.14 $\\
$56513.6580$ & $-85.93$ & $\pm	0.15$ & $51.77$ & $\pm	0.14 $\\
$56513.6567$ & $-86.56$ & $\pm	0.14$ & $51.52$ & $\pm	0.17 $\\
$56515.6546$ & $47.53$ & $\pm	0.13$ & $-85.76$ & $\pm	0.14 $\\
$56515.6555$ & $47.40$ & $\pm	0.14$ & $-85.65$ & $\pm	0.13 $\\
$56515.6564$ & $47.32$ & $\pm	0.12$ & $-85.46$ & $\pm	0.14 $\\
$56515.6572$ & $47.31$ & $\pm	0.14$ & $-85.36$ & $\pm	0.14 $\\
$56515.6559$ & $47.03$ & $\pm	0.13$ & $-85.90$ & $\pm	0.12 $\\
$57438.3754$ & $44.14$ & $\pm	0.15$ & $-81.32$ & $\pm	0.14 $\\
$57438.4363$ & $35.53$ & $\pm	0.12$ & $-72.00$ & $\pm	0.13 $\\
$57438.5453$ & $17.94$ & $\pm	0.16$ & $-54.39$ & $\pm	0.13 $\\
$56955.6077$ & $-12.07$ & $\pm	0.19$ & $-24.00$ & $\pm	0.21 $\\
$56955.6088$ & $-12.72$ & $\pm	0.26$ & $-23.28$ & $\pm	0.33 $\\
$56955.6098$ & $-12.70$ & $\pm	0.24$ & $-23.24$ & $\pm	0.27 $\\
$56955.6109$ & $-12.76$ & $\pm	0.28$ & $-23.21$ & $\pm	0.27 $\\
$56955.6093$ & $-12.13$ & $\pm	0.26$ & $-23.31$ & $\pm	0.29 $\\
\hline
    \end{tabular}
    \tablefoot{In addition, we list the formal uncertainties, as produced by the TODCOR algorithm.}
    \label{tab:narval}. 
\end{table}
\newpage
\section{Observation log}
We present the log of our observations in Table \ref{tab:obslog}, along with the telescope configurations and calibrators used.
\begin{table*}[!t]
    \centering
    \caption{SPICA, MIRC--X and MYSTIC observation log}
    \begin{tabular}{c|cccl}\hline\hline
Date	&	MJD	&	Config.	&	Ins.	&	\multicolumn{1}{c}{Calibrators}						\\\hline
2023-10-15	& $60232.4456$ &	5T (E1)	&	S	&	HD\,222173, HD\,6658, HD\,20418, HD\,22192, HD\,39283, HD\,58142	\\	
	& $ $&		&	MM	&	HD\,39283, HD\,58142					\\	
	& $60232.5137$ &	5T (E1)	&	S	&	HD\,222173, HD\,6658 , HD\,20418, HD\,22192, HD\,39283, HD\,58142	\\	
	& $ $&		&	MM	&	HD\,39283, HD\,58142					\\	\hline
2023-10-17	& $60234.5351$ &	6T	&	MMS	&	HD\,50973						\\	\hline
2023-11-11	& $60259.3913$ &	6T	&	S	&	HD\,222173, HD\,58142					\\	
	& $ $&		&	MM	&	HD\,222173, HD\,58142, HD\,76644				\\	\hline
2023-11-13	& $60261.3785$ &	5T (S1)	&	S	&	HD\,12303, HD\,20677, HD\,24760, HD\,58142			\\	
	&  $ $&		&	Mi	&	HD\,12303, HD\,20677, HD\,58142				\\	\hline
2024-10-14	& $60597.4903$ &	6T	&	S	&	HD\,6961 , HD\,6658 , HD\,10205, HD\,32630			\\	
	& $ $&		&	MM	&	HD\,198639, HD\,6961, HD\,32630				\\	\hline
2024-10-31	& $60614.4209$ &	6T	&	S	&	HD\,25642, HD\,58142					\\	
	& $ $&		&	Mi	&	HD\,222173, HD\,25642, HD\,58142, HD\,76644			\\	
	& $ $&		&	My	&	HD\,222173, HD\,13041, HD\,20677, HD\,25642, HD\,58142, HD\,76644	\\	
	& $60614.4695$ &	6T	&	S	&	HD\,25642, HD\,58142					\\	
	& $ $&		&	Mi	&	HD\,222173, HD\,25642, HD\,58142, HD\,76644			\\	
	& $ $&		&	My	&	HD\,222173, HD\,13041, HD\,20677, HD\,25642, HD\,58142, HD\,76644	\\	\hline
2025-10-28	& $60976.4428$ &	6T	&	MM	&	HD\,3333 , HD\,12772, HD\,15139, HD\,18927, HD\,40205, HD\,47726	\\
\hline
    \end{tabular}
    \tablefoot{We list the individual calibrators used to obtain the transfer function. The Config. column describes configuration, i.e. the number of telescopes used, where in the case of 5-telescope observation (5T), the missing telescope is listed in parentheses. The Ins. column denotes the instruments performing the observation S -- SPICA, MM -- MIRC--X\&MYSTIC or, in the cases where a distinction is necessary, Mi refers to MIRC--X, while My to MYSTIC. On the night of Nov 13 2023, only MIRC--X and SPICA data were recovered.}
    \label{tab:obslog}
\end{table*}
\FloatBarrier
\section{Derivation of the profile likelihood}
\label{app:profile_likelihood}
To derive the equation of profile likelihood, we start from a Gaussian likelihood function with dataset-dependent weights $w_k$
\begin{equation}
    \ln \mathcal{L} = - \sum_k \frac{1}{2} w_k \chi_k^2 + \sum_k \frac{N_k}{2} \ln w_k,
    \label{eq:weighted_likelihood}
\end{equation}
where $k$ is the index of the observable and $N_k$ the number of data points in each set. Treating $w_k$ as a nuisance parameter, we can find the maximum likelihood with respect to the individual $w_k$
\begin{equation}
    0 = \frac{\partial}{\partial w_k} \ln \mathcal{L} = -\frac{1}{2}\chi^2_k + \frac{N_k}{2 w_k}.
\end{equation}
Solving for $w_k$ gives
\begin{equation}
    w_k = \frac{N_k}{\chi^2_k}
\end{equation}
and substituting back into Eq. \ref{eq:weighted_likelihood}
\begin{equation}
    \ln \mathcal{L} = - \frac{1}{2}\sum_k N_k + \sum_k \frac{N_k}{2} \left[\ln N_k - \ln \chi^2_k \right].
\end{equation}
Since $N_k$ remain constant, terms independent of the model parameters (i.e. independent of $\chi^2$) can be omitted without affecting the parameter estimation and we can simplify this equation to
\begin{equation}
    \ln \mathcal{L} = - \sum_k \frac{N_k}{2} \ln \chi^2_k,
\end{equation}
which is exactly the form used in Eq.~\ref{eq:profile_likelihood}.
\FloatBarrier
\newpage
\section{Corner and trace plot of our solution}
\begin{figure*}[!t]
    \centering
    \includegraphics[width=0.85\linewidth]{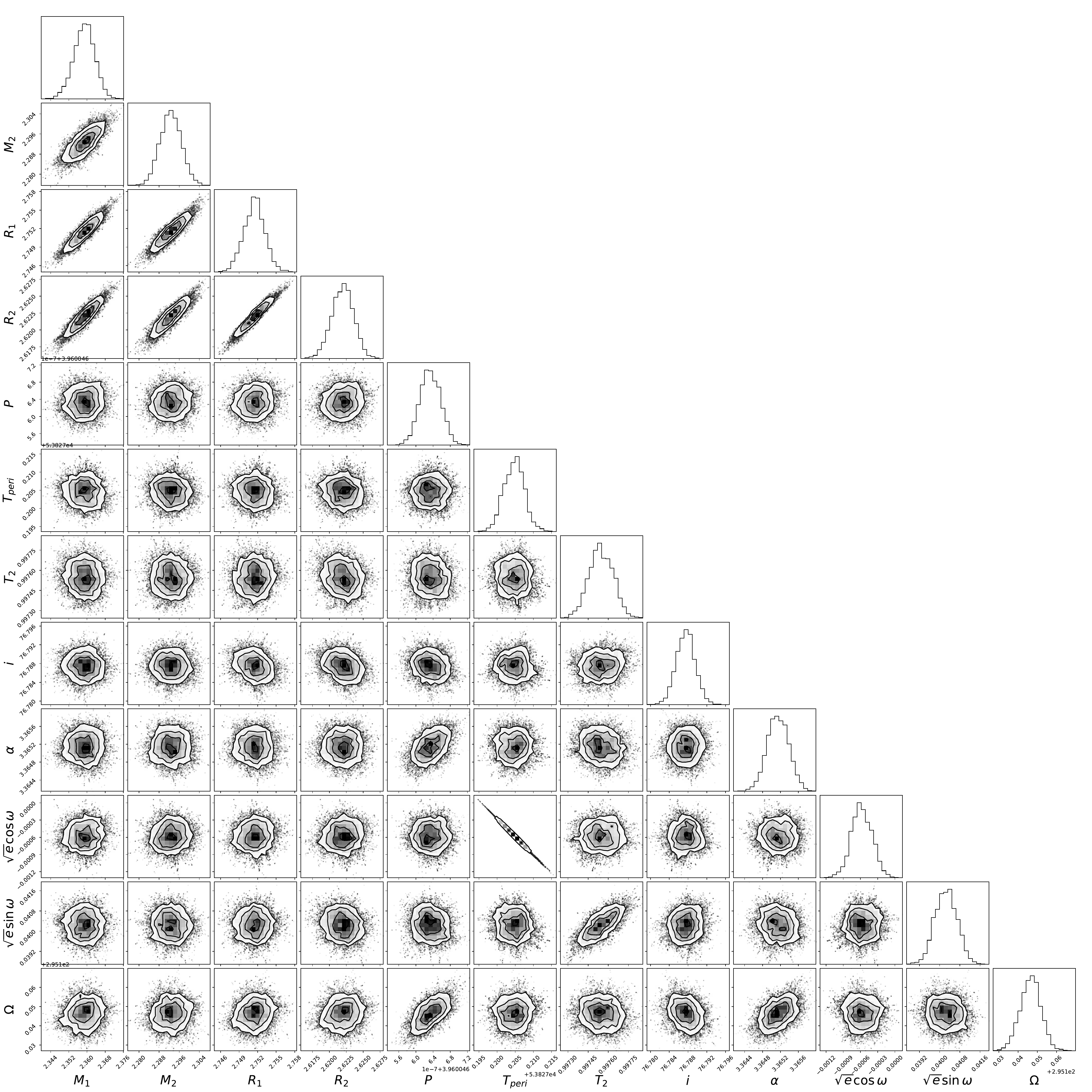}
    \caption{Corner plot of our MCMC solution. The unit of each parameter corresponds to the one listed in Table \ref{tab:result}.}
    \label{fig:solution}
\end{figure*}

\begin{figure*}[!t]
    \centering
    \includegraphics[width=0.85\linewidth]{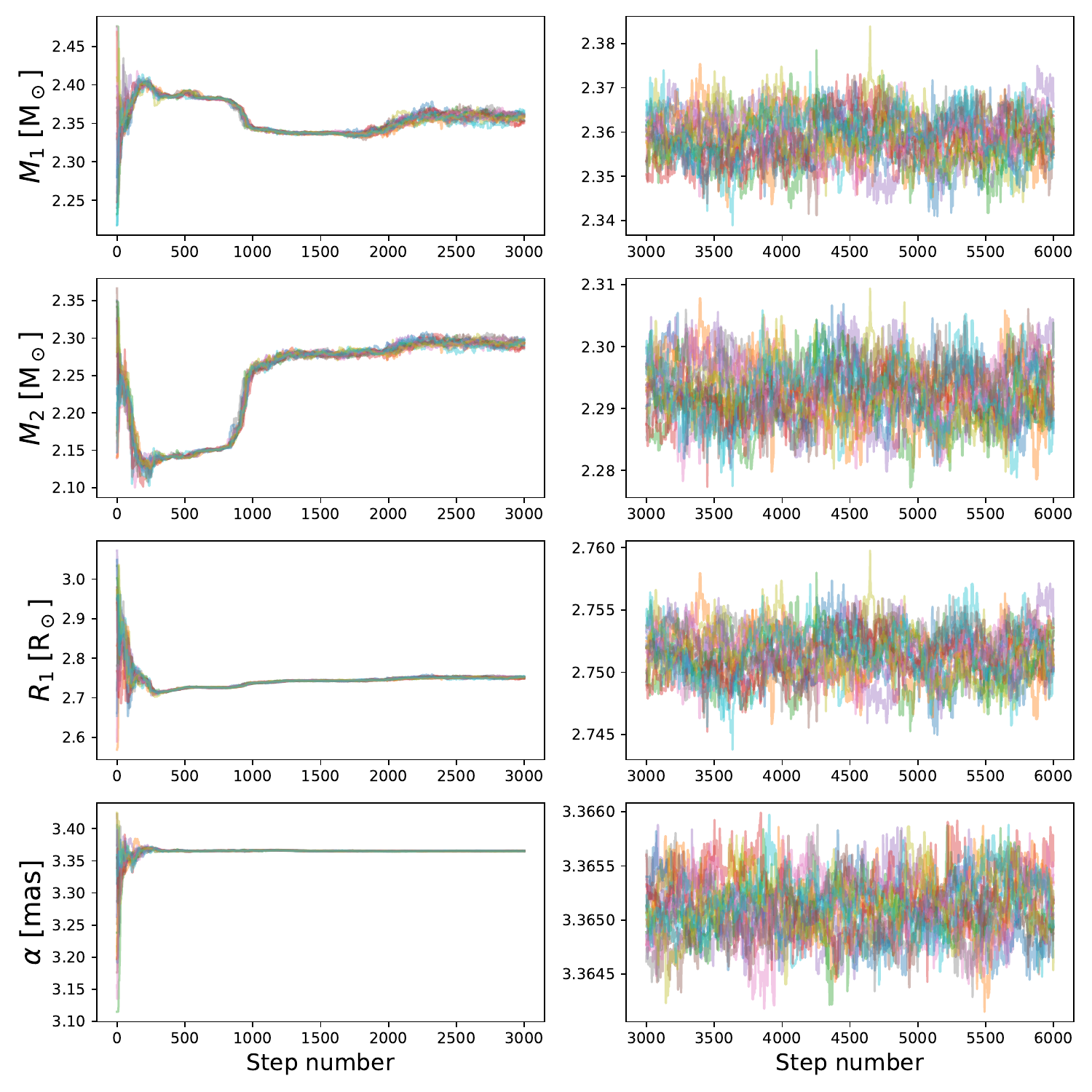}
    \caption{Trace plot of the convergence of the MCMC solution for four parameters, the masses $M_1$, $M_2$, the radius of the primary $R_1$ and the angular semi-major axis $\alpha$. The plot is separated to two parts, the initial burn-in phase (roughly 2500 steps) and the sampling phase (last 3000 steps).}
    \label{fig:trace}
\end{figure*}

In Fig. \ref{fig:solution}, we display the resulting corner plot of our solution, and in Fig. \ref{fig:trace} the evolution of the walkers for four of the individual parameters.


\end{appendix}

\end{document}